\documentclass[twocolumn,showpacs,preprintnumbers,amsmath,amssymb,prl]{revtex4}

\usepackage{graphicx}
\usepackage{dcolumn}
\usepackage{bm}
\usepackage{amsmath}
\usepackage{float}
\usepackage{siunitx}
\usepackage{textcomp}
\usepackage{comment}
\usepackage{xcolor}

\begin{document}


\title{Magnetosensitivity in dipolarly-coupled three-spin systems}

\author{Robert H. Keens}
\author{Salil Bedkihal}%
\author{Daniel R. Kattnig}
 \email{D.R.Kattnig@exeter.ac.uk} 
\affiliation{%
Living Systems Institute and Department of Physics, University of Exeter, Stocker Road, Exeter, Devon, EX4 4QD, United Kingdom
 }%


\date{\today}

\begin{abstract}
The Radical Pair Mechanism is a canonical model for the magnetosensitivity of chemical reaction processes. The key ingredient of this model is the hyperfine interaction that induces a coherent mixing of singlet and triplet electron spin states in pairs of radicals, thereby facilitating magnetic field effects (MFEs) on reaction yields through spin-selective reaction channels. We show that the hyperfine interaction is not a categorical requirement to realize the sensitivity of radical reactions to weak magnetic fields. We propose that, in systems comprising three instead of two radicals, dipolar interactions provide an alternative pathway for MFEs. By considering the role of symmetries and energy level crossings, we present a model that demonstrates a directional sensitivity to fields weaker than the geomagnetic field and remarkable spikes in the reaction yield as a function of the magnetic field intensity; these effects can moreover be tuned by the exchange interaction. Our results further the current understanding of the effects of weak magnetic fields on chemical reactions, could pave the way to a clearer understanding of the mysteries of magnetoreception and other biological MFEs and motivate the design of quantum sensors. Further still, this phenomenon will affect spin systems used in quantum information processing in the solid state and may also be applicable to spintronics.
\end{abstract}

\maketitle

There is growing excitement about the possibility of quantum coherence and entanglement underpinning the optimal functioning of biological processes \cite{Lambert2013}. A notable example is the avian inclination compass  \cite{Hore2016,Hiscock26042016, PhysRevLett.106.040503,NoMoreKominis,PhysRevE.87.062704,PhysRevLett.109.220501,PhysRevA.85.022315,RITZ, Tiersch4517,Ming2013,Daniel2016NJP}, which has recently been realised as a truly quantum-biological process  \cite{Hiscock26042016}. The leading explanation of this phenomenon utilizes the Radical Pair Mechanism (RPM), which describes the unitary evolution of singlet-triplet (S-T) coherences in systems comprising two radicals, i.e. two electron spins \cite{Hore2016,Rodgers353,Steiner1989}. The RPM has also been suggested to underpin controversial health-related implications of exposure to weak electromagnetic fields \cite{ghodbane2013bioeffects, Lalo1994, Kabuto2001, Juutilainen2018}. For these phenomena, the so-called €˜low-field effect€™ (LFE) is crucial to foster sensitivity to magnetic fields of intensity comparable to the geomagnetic field ($\approx 50~\mu${T}) \cite{Timmel1998, Maeda2008,Maeda2012,Suzuki2005,PhysRevE.87.062704,Kattnig2016_1, doi:10.1021/acs.jpcb.7b07672}.
The electron-electron dipolar interaction is often neglected when addressing MFEs within the RPM framework, but preliminary explorations have been conducted: electron-electron dipolar coupling is expected to resemble the exchange coupling, which, as the dominant interaction, suppresses S-T conversion by lifting the near-degeneracy of triplet and singlet states, reducing their susceptibility to mixing by weak hyperfine interactions \cite{doi:10.1021/jp5039283}, and quenching the LFE \cite{doi:10.1021/jp0456943}. Efimova et al. proposed that the dipolar interaction could be partly compensated by the exchange interaction, thereby allowing high sensitivity to the geomagnetic field despite sizeable electron-electron dipolar coupling interactions \cite{BiophysicalHore2008}.   

In contrast to the two-spin systems of the classical RPM, spin triads have attracted comparably little attention. Systems of three spins have been discussed: in the context of spin catalysis \cite{Buchachenko1996}, the chemical Zeno effect \cite{Letuta2015, doi:10.1021/acs.jpcb.7b07672}, quantum teleportation \cite{Salikhov2007}, and as a decoherence pathway \cite{Molin2013}. In spin catalysis, the exchange coupling of the radical pair with the spin catalyst is the main interaction motif. As the Zeeman part of the Hamiltonian commutes with the exchange Hamiltonian, this interaction alone is insufficient to produce MFEs (see SI \cite{SI}). However, mutual exchange coupling can provide the premise for near level-crossings at certain strengths of an external magnetic field, whereupon hyperfine-driven spin conversion can proceed efficiently \cite{magin2004, magin2005}, and may also transmit the effect of a fast-relaxing third radical \cite{HorePsynth}. A perturbative approach based on a Hubbard-trimer Hamiltonian has been used to show that the additional radical can enhance the intersystem crossing rate \cite{doi:10.1021/ja807590q}. Spin coherence transfer in the three-radical system has been recently realized experimentally \cite{doi:10.1021/ja807590q}. Furthermore, the spin-selective reaction of a radical pair with a scavenger radical has been shown, boost anisotropic magnetic field effects \cite{kattnig2017sensitivity} and provide resilience to spin relaxation in one of the radicals of the triad \cite{doi:10.1021/acs.jpcb.7b07672}, thereby providing decisive advantages over the classical RPM model of magnetoreception. To the authors' knowledge, three-spin systems have only been discussed in the biological context in \cite{Salikhov2007, doi:10.1021/acs.jpcb.7b07672, kattnig2017sensitivity}. All models mentioned in this context have disregarded the effects of electron-electron dipolar interactions.

We consider a toy model of three spins in an external magnetic field and investigate the MFEs that arise as a consequence of inter-radical interaction. 
Our model Hamiltonian is given (in angular frequency units) by
\begin{equation}
\label{eq1}
\begin{aligned}
\hat{H} &=\hat{H}_{0}+\hat{H}_{1}=\hat{H}_{dd}+\hat{H}_{ex}+\hat{H}_{1} \\
  &= \resizebox{0.75\hsize}{!}{$\sum_{i<j}^N{\bf{\hat{S}}}_{i}\cdot{\bf{D}}_{i,j}\cdot{\bf{\hat{S}}}_{j}-\sum_{i<j}^NJ_{i,j}\left(\frac{1}{2}+2{\bf{\hat{S}}}_{i}\cdot{\bf{\hat{S}}}_{j}\right)+ \gamma\roarrow{\bf{B}_{0}}\cdot\sum_{i}^N{\bf{\hat{S}}}_{i}$}
\end{aligned}
\end{equation}
The individual summands account for the electron-electron dipolar ($\hat{H}_{dd}$), exchange ($\hat{H}_{ex}$) and Zeeman interactions ($\hat{H}_{1}$). $\roarrow{\bf{B}_0}$ denotes the applied magnetic field, $B_0$ its intensity and $\gamma = \frac{g\mu_B}{\hbar}$. 
Here, we have assumed that the Zeeman interaction is isotropic and identical for all radicals on account of our focus on the MFEs of organic radicals in weak magnetic fields, i.e. $g\approx2$ and the anisotropies are negligible for moderate $B_{0}$. The electron-electron dipolar interactions are treated in the point-dipole limit. The interaction energy is related to the (supra)-molecular structure of the spin-triad by 
\begin{equation}
 \resizebox{0.9\hsize}{!}{${{\bf{\hat{S}}}_{i}\cdot{\bf{D}}_{i,j}\cdot{\bf{\hat{S}}}_{j}}=d_{i,j}(r_{i,j})\left[{{\bf{\hat{S}}}_{i}\cdot{\bf{\hat{S}}}_{j}}-3\left({\bf{\hat{S}}}_{i}\cdot\roarrow{\bf{e}}_{i,j}\right)\left({\bf{\hat{S}}}_{j}\cdot\roarrow{\bf{e}}_{i,j}\right)\right].$}
\end{equation} 
In the above equation, $\roarrow{\bf{e}}_{i,j} = \frac{\roarrow{\bf{r}}_{i,j}}{\lvert\roarrow{\bf{r}}_{i,j}\rvert}$ where $\roarrow{\bf{r}}_{i,j}$  is the vector connecting radical centres $i$ and $j$, $d_{i,j} = \frac{d_{0}}{\lvert\roarrow{\bf{r}}_{i,j}\rvert^{3}}$, and $d_{0} = \frac{\mu_{0}}{4\pi\hbar}g^{2}\mu_{B}^2$.
We assume that the three-radical system is generated in the singlet state of radicals $1$ and $2$ with the third radical uncorrelated to the others, i.e. the initial density operator obeys $\hat{\rho}(t=0)=\frac{1}{2}\hat{P}_{s}^{(1,2)}$ where $\hat{P}_{s}^{(i,j)}=\frac{1}{4}-{\bf{\hat{S}}}_{i}\cdot{\bf{\hat{S}}}_{j}=\frac{1}{2}(1-\hat{P}_{i,j})$ is the singlet projection operator on the $i,j$- subspace, and $\hat{P}_{i,j}$ is the permutation operator for spins $i$ and $j$.
Assuming that radicals $1$ and $2$ recombine with equal rate constant $k$ in the singlet and triplet configurations, the equation of motion for the spin-triad density matrix becomes 
\begin{equation}
\frac{d\hat{\rho}}{dt}=-i[\hat{H}, \hat{\rho}(t)]-k\hat{\rho}(t).
\end{equation}
The quantum yield of the singlet recombination product of radicals $1$ and $2$ is $\varphi_{s}=k\int_{0}^{\infty}d\tau Tr[{\hat{P}_{s}}^{(1,2)}\hat{\rho}(\tau)]$, and the powder-averaged singlet yield is $\langle \varphi_{s}\rangle=\frac{1}{4\pi}\int_{0}^{2\pi}d\phi\int_{0}^{\pi}d\vartheta \sin({\vartheta})\varphi_{s}(B_{0}(\vartheta, \phi))$.
The MFEs can then be quantified by $\chi_{s}=\frac{\varphi_{s}(B_{0})}{\varphi_{s}(0)}-1$; analogous definitions apply to the orientation-averaged yield. 
In the eigen-basis of the Hamiltonian, $\hat{H}$, we find that 
$\varphi_{s}=\frac{1}{2}\sum_{i,j}{\left | \langle i|P_{s}^{(1,2)}|j\rangle\right |}^{2} f(k, \omega_{i}-\omega_{j}),$
where $f(k,\Delta\omega)=\frac{k^2}{k^2+{\Delta\omega}^{2}}$, and $\lvert i\rangle$, $\lvert j\rangle$ are the eigenstates of $\hat{H}$.
We show in the SI that the conclusions we draw using this approach are still qualitatively valid if recombination proceeds at different rates in the singlet and triplet configuration. \par
\begin{figure}[ht!]
\centering
\includegraphics[width=85mm]{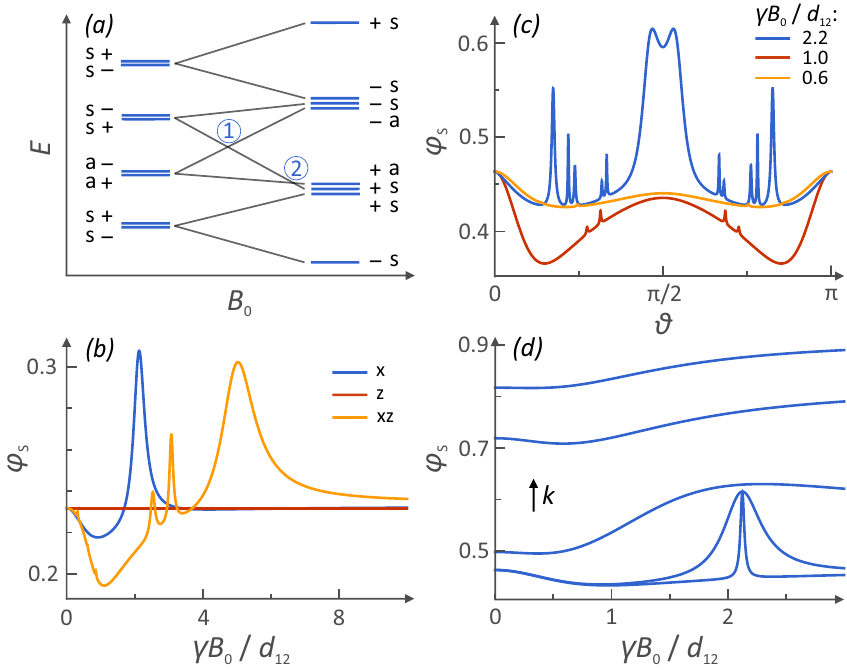}
\caption{(a) Schematic correlation diagram of energy level crossings as a function of the applied field, $B_{0}$. The labels classify the (anti-)symmetry of the states under $\hat{P}_{1,3}$ and $\hat{X}$. (b) Yield vs orientation for selected values of $B_{0}$, here, the recombination constant $k/d_{1,2} = 0.01$. (c) Yield vs $B_{0}$ for selected orientations; here $k/d_{1,2} = 0.01$. (d) Yield vs $B_0$ for $k/d_{1,2} = 0.001, 0.01, 0.1, 0.5, 1$, respectively, and in ascending order. Here the Zeeman field is along the $x$-direction. In all of the above figures, $J = 0$.}
\label{Correlation}
\end{figure}
The magnetic field independent part of the Hamiltonian ($\hat{H}_{0}$) is invariant under time reversal symmetry, i.e it commutes with the time reversal operator $\hat{\Theta}=e^{i\pi\hat{S}_{y}}\hat{\mathcal{K}}$ where $\hat{\mathcal{K}}$ denotes complex conjugation in the standard basis, and $\hat{S}_{y}$ is the $y$-component of the total spin-angular momentum operator $\hat{\mathbf{S}}=\sum_{j}\hat{\mathbf{S}}_{j}$. As $\hat{\Theta}^{2}=-1$, the eigenstates of $\hat{H}_{0}$ are (at least) two-fold degenerate (Kramers degeneracy \cite{klein1952}). Furthermore, as $\hat{\Theta}$ maps $\lvert S^{(1,2)} \pm\rangle$ into $\pm\lvert S^{(1,2)} \mp\rangle$, pairwise degenerate states ($\lvert i\rangle$, $\hat{\Theta}\lvert i\rangle$) yield the same expectation value of ${\hat{P}_{s}}^{(1,2)}$. Note however, that $\lvert S^{(1,2)} \pm\rangle$ is not an eigenstate of the Hamiltonian, in stark contrast to the well-studied scenarios of pairs of radicals. 
This Kramers degeneracy, in spin triads, is broken by an external magnetic field. Consequently, the energy levels split and, depending on symmetry properties, can cross and/or anti-cross as a function of the applied field. This gives rise to prominent MFEs by impacting upon the coupling matrix elements $\langle i\lvert{\hat{P}_{s}}^{(1,2)}\rvert j\rangle$ and $f(k, \Delta\omega)$ (through altered energy differences). An example of such degeneracy-lifting is shown schematically in Fig.~\ref{Correlation}(a). The necessary and sufficient conditions to observe MFEs are that (i) $[\hat{P}_{s}^{(1,2)},\hat{H_{0}}]\neq 0$ and (ii) $\hat{H}_0$ does not possess the $SU(2)$ spin rotation symmetry; see SI for details. For a radical \textit{pair}, $\hat{H}$ always commutes with ${\hat{P}_{s}}^{(1,2)}$, and no MFEs are observed due to inter-radical interactions. Here, we argue that dipolarly-coupled spin triads give rise to MFEs for all configurations except for a peculiar one with the third (inert) radical placed halfway between the recombining radicals (and in the limit that radical $3$ is so remote that it does not impact upon the spin-evolution of the dyad on the timescale of its lifetime; see SI.  
On the contrary, a purely exchange-coupled isotropic spin system 
does not exhibit magnetosensitivity as a consequence of the retained $SU(2)$ symmetry of $\hat{H}_0$, as shown in the SI \cite{SI}. 

We first considered a linear symmetric triad for which $d_{1,2} = d_{2,3} = 8~d_{1,3}$; the effect of changes to this geometry can be seen below and, in more detail, in the SI \cite{SI}. 
For negligible exchange couplings ($J_{i,j} = 0$), Figs.~\ref{Correlation} (b-d) show the singlet yield as a function of the magnetic field for selected orientations or as a function of orientation for selected fields.
 In the case where the magnetic field is parallel to the molecular symmetry axis, $z$, $[\hat{H}_{dd},\hat{H}_{1}] = 0$, i.e. the $z$-component of the total magnetization is conserved and no MFE arises. An analytic calculation reveals the dependence of the singlet yield on $k$ \cite{SI}. In the limit of slow recombination, the yield approaches $\frac{841}{1816}$; for fast recombination no spin conversion is observed ($\varphi_{S} = 1$) as expected. For any other orientation of the magnetic field, the Zeeman Hamiltonian does not commute with the dipolar part and pronounced MFEs can be observed, as demonstrated in Fig.~\ref{Correlation}(b). With $B_{0}||x$, a marked spike is observed for $B_{0} \approx 2.1~ d_{1,2}$. This peak is the consequence of the crossing of two energy levels, with different permutation symmetries but the same spin-inversion symmetry, as is schematically illustrated in the correlation diagram in Fig.~\ref{Correlation}(a). $\hat{H}$ is symmetric with respect to the interchange of spins $1$ and $3$. Consequently, six of the eigenstates are symmetric and two are anti-symmetric with respect to $\hat{P}_{1,3}$.
For $B_{0}||x$, the latter two are proportional to $\lvert \alpha\alpha\beta\rangle\pm\lvert\alpha\beta\beta\rangle-\lvert\beta\alpha\alpha\rangle\mp\lvert\beta\beta\alpha\rangle$. The second of these (lower sign) crosses two of the states of the symmetric representation at $B_{0}\approx 0.3812~d_{1,2}$ and $B_{0}=\frac{17}{8}~d_{1,2}$.
The spike results from the second of these crossings (labelled $(2)$ in Fig. \ref{Correlation}(a)); for the first one, the matrix element of $\langle i\lvert{\hat{P}_{s}}^{(1,2)}\rvert j\rangle$ vanishes by symmetry. This is the case because, for $B_{0}||x$, $\hat{X}=\otimes_{i}\hat{\sigma}_{i,x}$, which exchanges $\alpha$ and $\beta$ states, provides another symmetry element. Since $\left[\hat{X}, \hat{P}_{s}^{(1,2)}\right]=0$, only crossings of the same $\hat{X}$-symmetry can alter the MFE. As such, sharp changes in the reaction yield result from the level crossings between states of different symmetry provided that the off-diagonal matrix elements of the singlet projection operator do not vanish. 
For arbitrary magnetic field and orientation, the singlet yield has to be evaluated numerically. For high fields, all but the secular parts of $\hat{H}_{dd}$ can be neglected and a perturbation-theoretical treatment yields an analytic expression of the singlet yield and its orientational dependence (given in the SI \cite{SI}). Its most obvious feature is the cessation of spin evolution for the magic angle.

\begin{figure}[!]
 \centering
  \includegraphics[height=6.8cm]{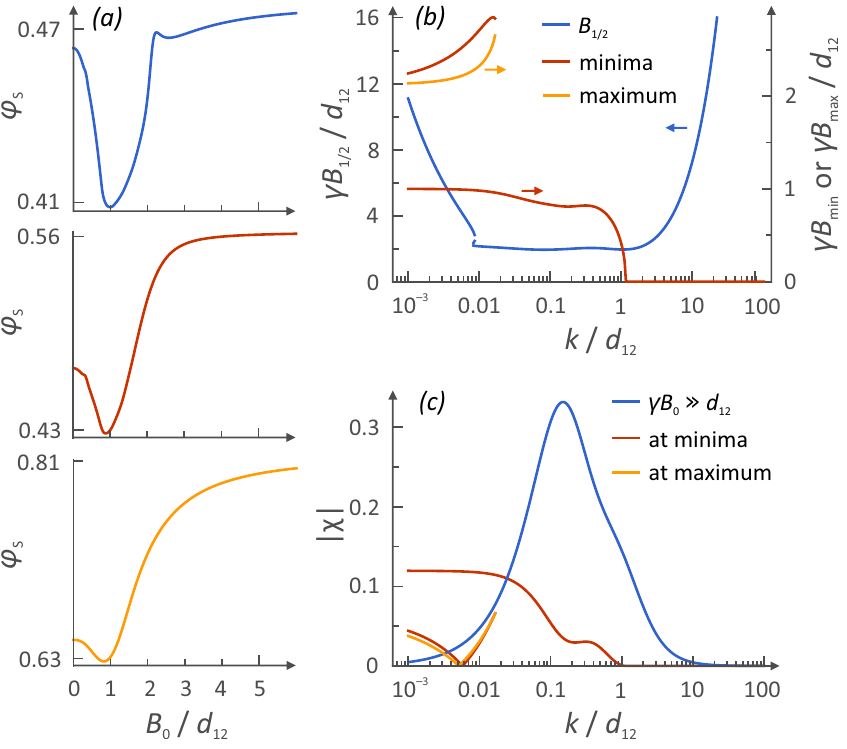}
 \caption{(a) Powder averages for $k/d_{1,2} = 6.86 \cdot 10^{-3}$, $0.047$ and $0.322$, respectively. (b) MFE characterized by half saturation $B_{\frac{1}{2}}$, and the locations of minima ($B_{min}$) and maxima ($B_{max}$), as a function of $k/d_{1,2}$. (c) Absolute MFE vs $k/d_{1,2}$. In all of the above figures $J = 0$.}
 \label{Powder}  
 \end{figure}

For a disordered system, the observed singlet yield represents the average over all possible orientations of the external magnetic field vector $B_{0}$, with respect to the molecular axes of the system. Fig.~\ref{Powder}(a) shows this powder average of the singlet yield for the linear, symmetric spin triad over a range of $k$-values. Interestingly, the recombination yield is not averaged to zero even though the average dipolar interaction of a pair of spins vanishes. The field-dependence is characterized by a minimum at $B_{0} \approx d_{1,2}$, which is the dominant feature at small $k$. We characterize the field-effect by established measures such as the field of half-saturation $B_{\frac{1}{2}}$ (the field for which $\langle \varphi_{s}(B_{0})\rangle$ equals $\frac{1}{2}(\langle \varphi_{s}(B=0)\rangle+\langle\varphi_{s}(B\to\infty)\rangle)$  and the MFEs associated with characteristic points such as the low-field minimum, which resembles the LFE documented for the hyperfine mechanism in radical pairs \cite{Timmel1998}. These parameters are summarized in Figs.~\ref{Powder}(b) and (c) as a function of $k$. For $B_{0}$ exceeding a few $d_{1,2}$, huge field effects $\chi$ in excess of $30~\%$, can be realized for intermediate $k$ of the order of $0.2~d_{1,2}$ (Fig.~\ref{Powder}(c)). The magnitude of the low-field feature is approximately $12~\%$  for small $k$-values. For context, note that at a distance of $17~$\AA, $d_{1,2}$ will be of the order of $10$~MHz and $k = 0.01~d_{1,2}$ would then correspond to a lifetime of $k^{-1} = 1.6~\mu$s. Under these conditions, the low-field effect in dipolarly-coupled spin triads is expected to closely agree with what the hyperfine mechanism can deliver for radical pairs (\textit{vide infra} for a substantial enhancement). Note that significant MFEs can also ensue for comparably short coherence times and, thus, quickly-relaxing radicals could be meaningfully considered in the triad model for small $r$.

\begin{figure}[ht]
\centering
\includegraphics[width=65mm]{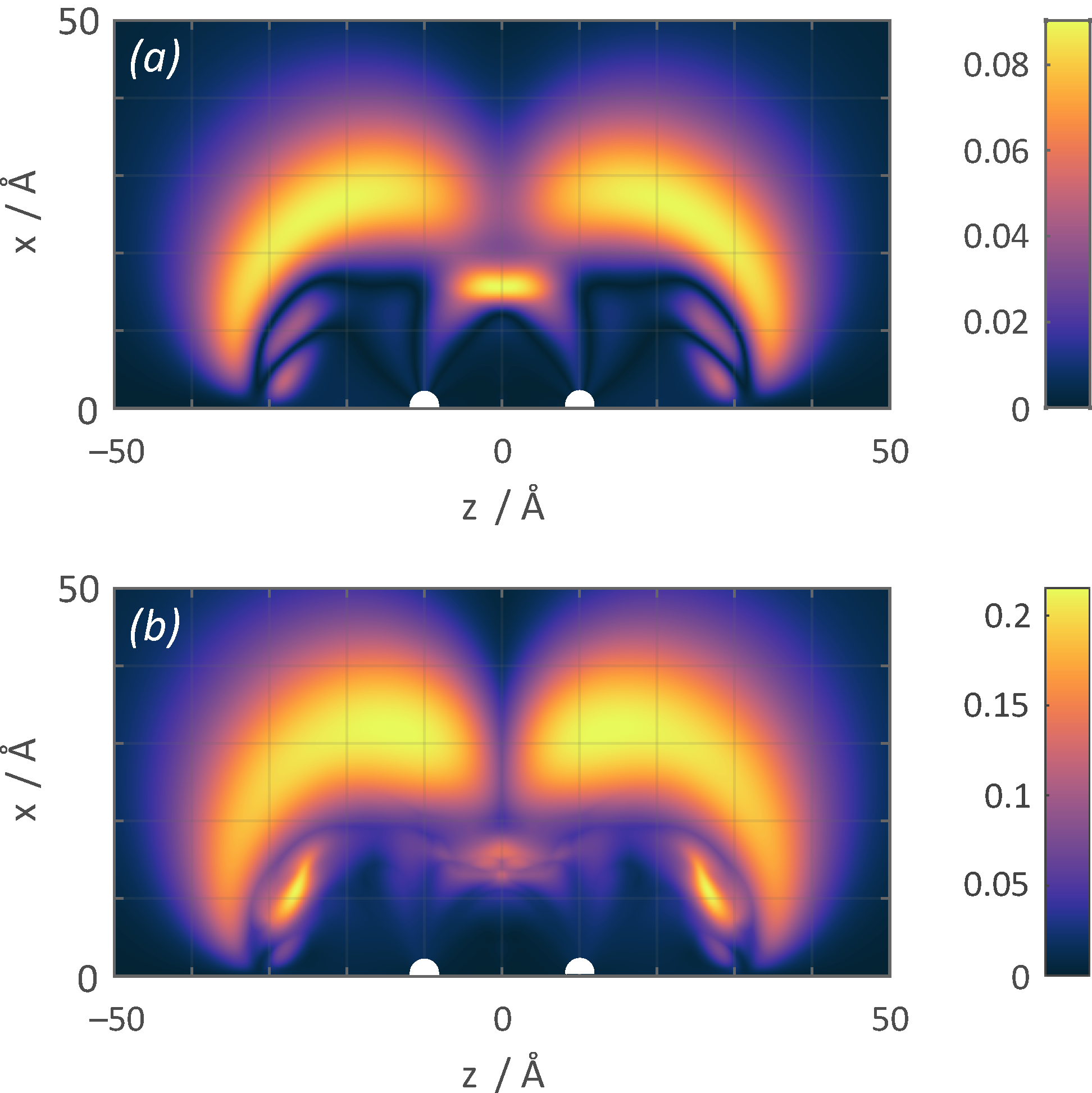}
\caption{(a) Absolute value of the MFE. (b) Anisotropy relative to the mean singlet yield. We use $k = 0.0245~d_{1,2}$ and $B_{0} = 0.215~d_{1,2}$ corresponding to a lifetime of $1~\mu${s} and the geomagnetic field ($50~\mu${T}) for an inter-radical distance of $20~$\AA, and $J = 0$. Spins $1$ and $2$ are located at $(0,a)$ and $(0, -€"a)$, respectively with $a = 10~$\AA, and the position of spin $3$ is varied in the containing plane.}
 \label{Anisotropy}
 \end{figure}

Substantial MFEs can in fact be observed for a variety of geometries of the spin triad. Fig.~\ref{Anisotropy} shows the dependence of the powder-averaged MFE and the relative anisotropy, i.e.\ the largest orientational spread of the singlet yield, relatively to the mean singlet yield, of general configurations for a magnetic field intensity of $50~ \mu${T} (roughly the geomagnetic field).
 Assuming that spins $1$ and $2$ are located at $(0,a)$ and $(0, -€"a)$, respectively, with $a = 10~$\AA, the maximum averaged MFE is $\approx 9~\%$ at the location of the third spin $(\pm1.58~a,~\pm{a})$. The maximum anisotropy amounts to $21.5~\%$ at $(\pm{3.18~a},~\pm{1.35~a})$, which corresponds to inter-radical distances as large as $40~$\AA. 
Thus sizeable MFEs are induced by the dipolar interaction even at relatively large distances, demonstrating that the effect does not rely on infrequent direct three-particle encounters, which could have a bearing on its relevance \cite{doi:10.1021/acs.jpcb.7b07672}. In the SI we show analogous results for $a = 7.5~$\AA~and higher field intensities, with anisotropies in excess of $100~\%$ and MFEs far above $30~\%$ \cite{SI}.

We further discuss the bond angle dependence for randomly oriented isosceles spin-triads. For the geomagnetic field we observe the largest field effect for bond angles roughly corresponding to a pentagon's internal angle ($144$\textdegree) or slightly less than an equilateral triangle's ($60$\textdegree). For greater field intensities, large effects can be realized for all bond angles of practical relevance (see SI). Unlike for the linear geometry, we find that the MFEs of these systems typically do not decrease with increasing $k^{-1}$. For the equilateral triangular geometry, sizeable MFEs for $k$ as large as $10~d_{1,2}$ are predicted. These observations indicate that geometry indeed plays an important role, and that the MFEs in certain geometries may be less susceptible to variations in the lifetime. The system typically shows avoided crossings of energy levels, which nonetheless can give rise to spiky features \cite{Hiscock26042016, sosnovsky2016level}.

\begin{figure}[ht]
 \centering
  \includegraphics[width=85mm]{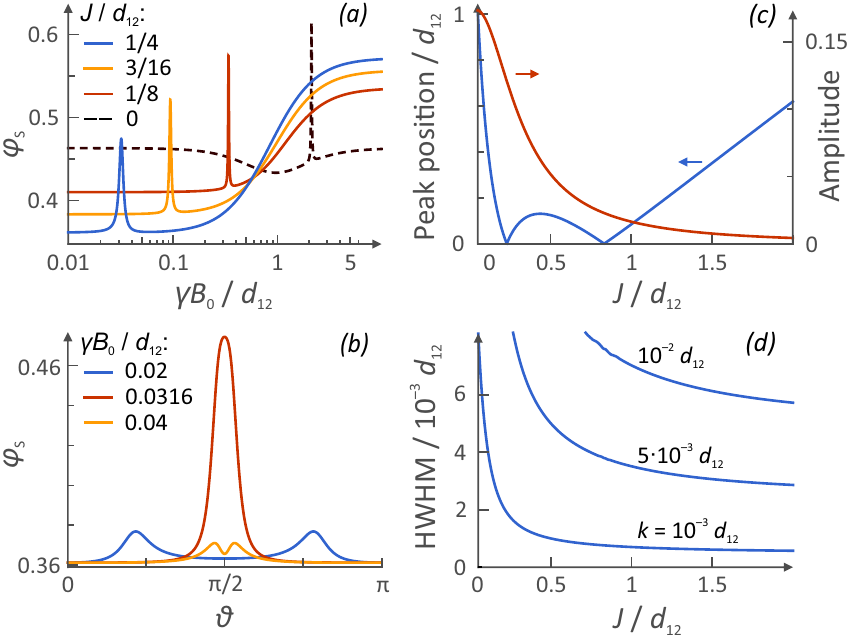}
 \caption{(a) Yield vs applied field for different exchange interaction strengths, $J/d_{1,2}$; here, $k/d_{1,2} = 0.001$ and $B_{0}||x$. (b) Yield as a function of orientation for different $B_{0}$; we use $k/d_{1,2} = 0.001$. (c) The location and amplitude of a low-field peak's maximum, as a function of the exchange interaction strength with $k/d_{1,2} = 0.001$ and $0.01$ (indistinguishable). (d) The half-width at half-maximum of a low-field peak as a function of $J/d_{1,2}$, for $k$ as indicated in the figure.}
 \label{Exchange} 
 \end{figure}
We further studied the effect of an additional exchange interaction on the MFEs. As $\hat{H}_{ex}$ displays time-reversal symmetry the eigenstates of the Hamiltonian are still double degenerate for $B_{0} = 0$. Yet, remarkable LFEs can emerge if the (anti-)crossings of energy levels are shifted to lower magnetic fields. An illustrative example of this phenomenon is provided by the linear spin-triad for $J = J_{1,2} = J_{2,3}$ and $J_{1,3} = 0$, i.e. for the symmetric coupling of adjacent spins, and with the magnetic field perpendicular to the triad axis. As shown in Fig. 4(a), the singlet yield of this system exhibits a sharp peak, which shifts to lower magnetic field intensities for exchange couplings approaching 0.25~$d_{1,2}$. Fig. 4(c) then shows how the amplitude and field-location of the peak vary as a function of the exchange coupling. It is interesting to note that, for typical dipolar coupling strengths, this feature may occur at field values smaller than the geomagnetic field. Formally, for these regions of maximal low-field sensitivity, the peak shifts from positive to negative magnetic field intensities. As shown in Figs.~\ref{Exchange}(c) and (d), the peak decreases in amplitude with increasing $J$ and broadens as the recombination rate constant increases. It remains prominent for $k$ up to $0.02~d_{1,2}$ which, for typical parameters, equates to lifetimes of the order of microseconds (but could be less for smaller inter-radical distances). The SI summarizes the dependence of these characteristic parameters of the MFE as a function of $k$ \cite{SI}. Importantly, the spike does appear in powder averages, suggesting that it could be relevant to MFEs in the randomly oriented samples implicated in biological radical reactions. Fig.~\ref{Exchange}(b) shows how a large directional anisotropy can occur in the low-field regime. Here, $J$ has been fixed to $\frac{1}{4}~d_{1,2}$ and one can see that the spike results from an applied field $B_{0} = 0.0316~d_{1,2}$, while comparatively little directional anisotropy can be seen for other low-field intensities. For negative $J$, and $J > 2~d_{1,2}$, the line shape does not exhibit a pronounced maximum; the main feature is a minimum at higher field. 
\\
\\
We have shown that remarkable MFEs can emerge in radical triads due to the dipolar interaction. This realization extends our current understanding of the magnetosensitivity of chemical reactions by providing an additional mechanistic pathway. Unlike the well-established RPM, this three-radical effect does not rely on hyperfine interactions or differing $g$-factors. Due to the slow decay of the dipolar interaction with distance, aspects of this mechanism could be unexpectedly relevant, e.g. in the context of magnetoreception, the adverse health effects putatively associated with electromagnetic field exposure, or for purposefully engineered sensing applications, coherent control, quantum information processing with spins in the solid state and potentially spintronics \cite{Liu2017, PhysRevB.93.220402, Ghirri2017}. In particular, the three-radical pathway could underlie the putative magnetosensitivity of lipid peroxidation, a process that follows a free-radical chain mechanism predominantly involving peroxyl radicals derived from polyunsaturated fatty acids as chain carriers devoid of dominant hyperfine interactions \cite{lipid, ghodbane2013bioeffects, Lalo1994, Kabuto2001}. In this context, it is remarkable that the effect can in principle provide MFEs of considerable amplitude and sensitivity to fields comparable to the geomagnetic field. For many radicals, the spin density is spread over several magnetic nuclei. In this case, S-T transitions will be induced by the hyperfine interaction \textit{and} the dipolar coupling among the radicals of the spin triad. It is surprising that this mechanism has so far remained unexplored.

\section*{Acknowledgements}
\begin{acknowledgments}
We would like to thank The Royal Society (RG170378), the EPSRC (grant no. EP/R021058/1) and NVIDIA (GPU Grant Program) for financial and in-kind support.
\end{acknowledgments}

\bibliography{apssamp_shortened_italics.bib}

\pagebreak
\widetext
\begin{center}
\textbf{\large Supplemental Materials: Magnetosensitivity in dipolarly-coupled spin-triads}
\end{center}
\setcounter{equation}{0}
\setcounter{figure}{0}
\setcounter{table}{0}
\setcounter{page}{1}
\makeatletter
\renewcommand{\theequation}{S\arabic{equation}}
\renewcommand{\thefigure}{S\arabic{figure}}
\renewcommand{\bibnumfmt}[1]{[S#1]}
\renewcommand{\citenumfont}[1]{S#1}


\section{S1. Introduction}
In the supporting information we present the justification of relevant claims made in the main text. 
First we show that magnetic field effects (MFEs) are absent for a system of two radicals coupled by electron dipole-dipole and/or exchange interactions.
We then show that MFEs are absent in an exchange coupled isotropic linear $N$-spin system on account of $SU(2)$ symmetry. We evaluate
the relevant commutators to demonstrate the $SU(2)$ symmetry of an isotropic, exchange coupled, linear spin system.
We further show that the SU(2) symmetry is broken when dipole-dipole interactions
are included. We obtain the necessary and sufficient conditions to realize magnetic field
effects, and show that the dipolar interaction mixes singlet-triplet states for most configurations of spin-triad; there exist only two special cases where no MFEs are observed. This observation remains valid even in the presence of exchange interactions.
Finally, we show MFEs for the equilateral triangular and general geometries, and the bond-angle dependence in isosceles triangular triads.

\section{S2. Properties of the field independent Hamiltonian} 
The magnetic field independent part of the Hamiltonian ($\hat{H}_{0}$) is invariant under time-reversal symmetry, i.e it commutes with the time reversal operator $\hat{\Theta}=e^{i\pi\hat{S}_{y}}\hat{\mathcal{K}}$, where $\hat{\mathcal{K}}$ denotes complex conjugation in the standard basis, and $\hat{S}_{y}$ is the $y$-component of the total spin-angular momentum operator $\hat{S}=\sum_{j}\hat{S}_{j}$. As $\hat{\Theta}^{2}=-1$, the eigenstates of $\hat{H}_{0}$ are (at least) two-fold degenerate (Kramer degeneracy). Furthermore, as $\hat{\Theta}$ maps $\lvert S^{(1,2)} \pm\rangle$ into $\pm\lvert S^{(1,2)} \mp\rangle$, pairwise degenerate states ($\lvert i\rangle$, $\hat{\Theta}\lvert i\rangle$) yield the same expectation value of ${\hat{P}_{s}}^{(1,2)}$. Here, $S^{(1,2)}$ in the state specification denotes the singlet state of spins $1$ and $2$, and the plus (minus) sign refers to spin-up (down) for the third spin. Note that, for spin triads, $\lvert S^{(1,2)} \pm\rangle$ is not in general an eigenstate of the Hamiltonian, in stark contrast to the well studied scenario of pairs of radicals. This paves the way to intriguing MFEs that are not reliant upon hyperfine interactions.

\section{S3. Necessary and sufficient conditions to realize MFEs in an N-spin system}
We first consider a situation where we have a system of two radicals coupled by dipole-dipole and exchange interactions. The Hamiltonian is given by 
\begin{flalign}
\hat{H} &= \hat{H}_0 + \hat{H}_1,\\
\nonumber\text{with:}\\
\nonumber\hat{H}_{0}&= \mathbf{\hat{S}}_{1}\cdot{\bf{D}}_{1,2}\cdot{\mathbf{\hat{S}}_2} - J_{1,2}\left(\frac{1}{2}+2{\bf{\hat{S}}}_{1}\cdot{\bf{\hat{S}}}_{2}\right),\\
\nonumber\text{and:}\\
\nonumber\hat{H_1} &=  \gamma\roarrow{\mathbf{B}}_{0}\cdot{\left(\mathbf{\hat{S}}_{1} +\mathbf{\hat{S}}_2\right)},
\end{flalign}
where $D_{1,2}$ is the dipole-dipole tensor coupling spins $1$ and $2$, $J_{1,2}$ is the exchange coupling strength and $B_0$ is the intensity of an applied magnetic field. Noting that among the commutators between the two-spin operators, only those that can be represented by $\left[ {2{{\hat S}_{1,i}}{{\hat S}_{2,k}},2{{\hat S}_{1,j}}{{\hat S}_{2,k}}} \right] = {\varepsilon _{i,j,m}}\,i\,{\hat S_{1,m}}{\hat S_{2,k}}$ 
 and $\left[ {2{{\hat S}_{1,k}}{{\hat S}_{2,i}},2{{\hat S}_{1,k}}{{\hat S}_{2,l}}} \right] = {\varepsilon _{i,l,m}}\,i\,{\hat S_{1,k}}{\hat S_{2,m}}$ are non-zero, with ${\varepsilon _{i,j,m}}$ denoting the Levi-Civita symbol and the summation over $m$ being implied, the singlet projection operator $\hat{P}_{S}^{(1,2)}=\frac{1}{4}-\hat{S}_{1}\cdot\hat{S}_{2}$ commutes with the field-independent Hamiltonian $\hat{H}_0$. Furthermore, as $\hat{H}_1$, and consequently $\hat{H}$, commutes with $\hat{P}_{S}^{(1,2)}$, the singlet state is a field-independent eigenstate of the Hamiltonian and no MFEs on the singlet yield are observed (in the absence of additional interactions).
\\
\\
Now we examine the necessary and sufficient conditions to realize MFEs in a system of $N>2$ spins. Let us first consider a purely exchange coupled isotropic spin system given by the following Hamiltonian,
\begin{equation}
\hat{H}_{0}=\sum_{i<j}^{N}J_{i,j}[\hat{S}_{i, x}\hat{S}_{j,x}+\hat{S}_{i,y}\hat{S}_{j,y}+\hat{S}_{i,z}\hat{S}_{j,z}].
\end{equation} 
Let us denote the total spin operator along the respective direction as $\hat{S}_{i}=\sum_{m}\hat{S}_{m,i}$ where $i~\in~\{x,y,z\}$.
We evaluate the following commutator
\begin{equation}
\begin{aligned}
\left[\hat{H}_{0},\hat{S}_{z}\right]
& =\sum_{i,j}{J_{i,j}}\sum_{m}[\hat{S}_{i,x}\hat{S}_{j,x}+\hat{S}_{i,y}\hat{S}_{j,y}+\hat{S}_{i,z}\hat{S}_{j,z},\hat{S}_{m,z}]\\
&=\sum_{i,j}{J_{i,j}}\sum_{m}\left([\hat{S}_{i,x}\hat{S}_{j,x}, \hat{S}_{m,z}]+[\hat{S}_{i,y}\hat{S}_{j,y}, \hat{S}_{m,z}]+[\hat{S}_{i,z}\hat{S}_{j,z}, \hat{S}_{m,z}]\right).
\end{aligned}
\end{equation}
The above commutator can be further reduced to 
\begin{equation}
\begin{aligned}
\left[\hat{H}_{0},\hat{S}_{z}\right]&=\sum_{i,j}^{N}{J_{i,j}}([\hat{S}_{i,x}\hat{S}_{j,x},\hat{S}_{i,z}]+[\hat{S}_{i,y}\hat{S}_{j,y},\hat{S}_{i,z}]+[\hat{S}_{i,z}\hat{S}_{j,z},\hat{S}_{i,z}]\\&+[\hat{S}_{i,x}\hat{S}_{j,x},\hat{S}_{j,z}]+[\hat{S}_{i,y}\hat{S}_{j,y},\hat{S}_{j,z}]+[\hat{S}_{i,z}\hat{S}_{j,z},\hat{S}_{j,z}]).
\end{aligned}
\end{equation}
The first term can be written as
\begin{equation}
\begin{aligned}
\left[\hat{S}_{i,x}\hat{S}_{j,x},\hat{S}_{i,z}\right]=-i\hat{S}_{i,y}\hat{S}_{j,x}.
\end{aligned}
\end{equation}
Likewise we can simplify the remaining commutators in Eq. (3) and obtain,
\begin{equation}
\begin{aligned}
\left[\hat{H}_{0},\hat{S}_{z}\right]&=\sum_{i<j}^{N}J_{i,j}([\hat{S}_{i,x}, \hat{S}_{i,z}]\hat{S}_{j,x}+[\hat{S}_{i,y}, \hat{S}_{i,z}]\hat{S}_{j,y}+
[\hat{S}_{i,z}, \hat{S}_{i,z}]\hat{S}_{j,z}\nonumber \\&+\hat{S}_{i,x}[\hat{S}_{j,x}, \hat{S}_{j,z}]+\hat{S}_{i,y}[\hat{S}_{j,y}, \hat{S}_{j,z}]+\hat{S}_{i,z}[[\hat{S}_{j,z}, \hat{S}_{j,z}])\nonumber \\& =i\sum_{i<j}^{N}J_{i,j}(-\hat{S}_{i,y}\hat{S}_{j,x}+\hat{S}_{i,x}\hat{S}_{i,y}+0+\hat{S}_{i,y}\hat{S}_{j,x}-\hat{S}_{i,x}\hat{S}_{j,y}+0)\nonumber \\& =0.
\end{aligned}
\end{equation}

Similarly we can show that $[\hat{H}_{0},\hat{S}_{x}]=[\hat{H}_{0},\hat{S}_{y}]=0$. This implies that an isotropic exchange coupled spin model has $SU(2)$ symmetry and the eigenstates can be arranged as $SU(2)$ multiplets. As $\hat{H}_1$ commutes with $\hat{P}_{S}^{(1,2)}$, we thus do not observe MFEs for this scenario.

Now we examine these commutators by adding dipole-dipole interactions of the form $\hat{H}_{0}=\sum_{i<j}{{\bf{\hat{S}}}_{i}\cdot{\bf{D}}_{i,j}\cdot{\bf{\hat{S}}}_{j}}$. In this case we find that $[\hat{H}_{0}, \hat{S}_{x}]\neq 0$, and MFEs are observed if $[\hat{P}_{S}^{(i,j)},\hat{H}_{0}]\neq 0$. Typically, $[\hat{H}_{0}, \hat{S}_{i}]\neq 0$, but we remark that this is in particular not true for a linear triad aligned along $x$. In the next section we discuss the geometric conditions that have to be fulfilled (see below for details) in order to observe MFEs in dipolarly coupled three-spin systems.

\section{S4. Conditions for magnetic field effects in dipolarly coupled three-spin systems}
We intend to show that the dipolar interaction induces magnetic field dependent singlet-triplet mixing in all spin configurations except for one special case.
Without loss of generality, we assume that the three spins are located at the origin, displaced along the $z$-axis (spin $1$) and along an arbitrary direction in the $x$,$z$-plane that is inclined with respect to the $z$-axis by an angle $\theta$ (spin $3$), respectively. In units of $d_{1,2}$, the dipolar coupling tensors are then given by
\begin{equation} D_{1,2}=
\begin{pmatrix}
1 & 0 & 0 \\
0 & 1 & 0\\
0 & 0 & -2
\end{pmatrix}
\end{equation}

\begin{equation} D_{2,3}=\frac{1}{r^{3}}
\begin{pmatrix}
\frac{[3\cos(2\theta)-1]}{2} & 0 & -3\cos(\theta)\sin(\theta)\\
0 & 1 & 0\\
-3\cos(\theta)\sin(\theta) & 0 & -\frac{[1+3\cos(2\theta)]}{2}
\end{pmatrix}
\end{equation}
and 
\begin{equation}
D_{1,3}=\frac{1}{f(r,\theta)^3}
\begin{pmatrix}
1-G_{1}(r,\theta) & 0 & G_{2}(r,\theta)\\
0 & 1 & 0\\
G_{2}(r,\theta) & 0 & G_{1}(r,\theta)
\end{pmatrix}
\end{equation}
where $f(r,\theta)=(1+r^{2}-2r\cos(\theta))^{\frac{1}{2}}$, $G_{1}(r,\theta)=2 - 3\left(\frac{r\sin\theta}{f(r, \theta)}\right)^2$, and $G_{2}(r,\theta)=\frac{3r(1-r\cos{\theta})\sin{\theta}}{f(r, \theta)^2}$, and we assume no stacked spins (i.e. we exclude the case $r = 1$ and $\theta = 0$).\\
With this, the dipolar Hamiltonian becomes
\begin{equation}
\hat{H}_{0}=\hat{S}_{1}D_{1,2}\hat{S}_{2}+\hat{S}_{2}D_{2,3}\hat{S}_{3}+\hat{S}_{1}D_{1,3}\hat{S}_{3}.
\end{equation}
We are interested in the commutator  $[\hat{P}_{S}^{(1,2)}, \hat{H}] = [\hat{P}_{S}^{(1,2)}, \hat{H}_{0}]$. Using the commutator identities of the form $[\hat{A}\hat{B}, \hat{C}\hat{D}]=\hat{A}[\hat{B},\hat{C}]\hat{D}+\hat{A}\hat{C}[\hat{B},\hat{D}]+[\hat{A},\hat{C}]\hat{D}\hat{B}+\hat{C}[\hat{A},\hat{D}]\hat{B}$ and using angular momentum algebra, we can write the above commutators in the form of triple products of spin operators,
\begin{equation}
\begin{aligned}
-i[\hat{P}_{S}^{(1,2)},\hat{H}_{0}]=&
\left[\frac{1}{r^3}-\frac{1}{f(r,\theta)^5}\right]\left(\hat{S}_{1,z}\hat{S}_{2,x}\hat{S}_{3,y}-\hat{S}_{1,x}\hat{S}_{2,z}\hat{S}_{3,y}\right)\\&+
\left[\lambda (r, \theta)\right](\hat{S}_{1,y}\hat{S}_{2,x}\hat{S}_{3,x}+\hat{S}_{1,z}\hat{S}_{2,y}\hat{S}_{3,z}\\&-\hat{S}_{1,x}\hat{S}_{2,y}\hat{S}_{3,x}-\hat{S}_{1,y}\hat{S}_{2,z}\hat{S}_{3,z})+...,
\end{aligned}
\end{equation}
where \begin{equation}
\lambda(r,\theta)=\frac{3\cos\theta\sin\theta}{r^3}-\frac{3r\sin{\theta}(r\cos\theta-1)}{f^5}.
\end{equation}
The sum on the right hand side of Eq.~(6) comprises a total of $10$ different triple products of the spin operators of radical 1, 2 and 3, which can be collected in $4$ groups of unique dependence on $r$ and $\theta$.
The spin evolution mixes singlet and triplet states unless the singlet states are the eigenstates of the Hamiltonian $\hat{H}_{0}$, which requires the commutator [$\hat{P}_{S}^{(1,2)},\hat{H}_{0}$] to vanish; we seek the conditions on $r$ and $\theta$ for which this occurs.\\
The first set of terms vanishes for $\theta=\pm\arccos\left(\frac{1}{2r}\right)$. Inserting this into the second term gives a non-zero contribution except for $r=1/2$, which stipulates that $\theta=0, \pi/2, \pi$ on account of the first condition. Indeed, only for $\theta=0$ do all terms in the commutator vanish simultaneously. Thus, we conclude that the dipolar interaction mixes singlet and triplet terms except for one peculiar configuration, for which the third (unreactive) radical is placed half-way between the recombining radicals. Obviously, this corresponds to a scenario that might be difficult to realize in practice. If we additionally take the exchange interaction into account, an analogous treatment reveals that the same products of spin operators appear in the commutator. The coefficients are, however, in part augmented by the difference of $J_{1,3}$ and $J_{2,3}$. A detailed analysis reveals that the commutator vanishes if the above conditions are fulfilled ($r=1/2$, $\theta=0$) and additionally $J_{1,3} = J_{2,3}$.
\\
We continue by showing that the dipolar interaction gives rise to a magnetic field dependence of the singlet recombination yield for most configurations of three spins.
The reaction yield varies with the intensity of magnetic field if $\hat{H}_{0}$ does not commutes with the singlet projection operator, as established above, and $\hat{H}_{0}$ does not commute with $\hat{H}_{1}$. We shall focus on this second condition for the dipolar Hamiltonian. The relevant commutator becomes
\begin{equation}
\begin{aligned}
-i[\hat{H}_{dd}, \hat{H}_{1}]=-3\omega_{0,x}(\hat{S}_{1,z}\hat{S}_{2,y}+\hat{S}_{1,y}\hat{S}_{2,z})\\+3\omega_{0,y}(\hat{S}_{1,z}\hat{S}_{2,x}+\hat{S}_{1,x}\hat{S}_{2,z})+...
\end{aligned}
\end{equation}
The right hand side of above equation contains $20$ summands, which are bilinear in spin operators; $\omega_{0,i}$, with $i~\in \{x,y,z\}$, denotes the Larmor precession frequency associated with field-component $i$, and $\roarrow{\mathbf{\omega}}_0 = \gamma \roarrow{\mathbf{B}}_0$. The form of Eq. (12) suggests that the commutator can only vanish if $\omega_{0,x}=\omega_{0,y}=0$. Assuming this condition holds we arrive at the simpler expression
\begin{equation}
\begin{aligned}
-i[\hat{H}_{dd}, \hat{H}_{1}]=\omega_{0,z}\frac{3\sin{\theta}\cos{\theta}}{r^{3}}(\hat{S}_{2,z}\hat{S}_{3,y}+\hat{S}_{2,y}\hat{S}_{3,z})\\
+\omega_{0,z}\frac{3r(r\cos{\theta}-1)\sin{\theta}}{f^5}(\hat{S}_{1,z}\hat{S}_{3,y}+\hat{S}_{1,y}\hat{S}_{3,z})\\
+\omega_{0,z}\frac{3\sin^{2}{\theta}}{r^{3}}(\hat{S}_{2,y}\hat{S}_{3,x}+\hat{S}_{2,x}\hat{S}_{3,y})\\
+\omega_{0,z}\frac{3r^{2}\sin^{2}{\theta}}{f^5}(\hat{S}_{1,y}\hat{S}_{3,x}+\hat{S}_{1,x}\hat{S}_{3,y}).
\end{aligned}
\end{equation}
We can see that all the terms in the above equations are proportional to $\sin{\theta}$ and vanish for $\theta=0$ or $\theta=\pi$. As a result the singlet yield is insensitive to the magnetic field for a linear configuration of three radicals (including asymmetric configuration) if the magnetic field is parallel to the distinguished axis. For these orientations, the Hamiltonian has $U(1)$ symmetry and the magnetization along the distinguished axis is a conservative quantity.
Conversely, except for this peculiar scenario and the configuration discussed above, MFEs are generally predicted. 

\section{S5. Analytical results for the linear geometry}

\begin{figure}
 \centering
  \includegraphics{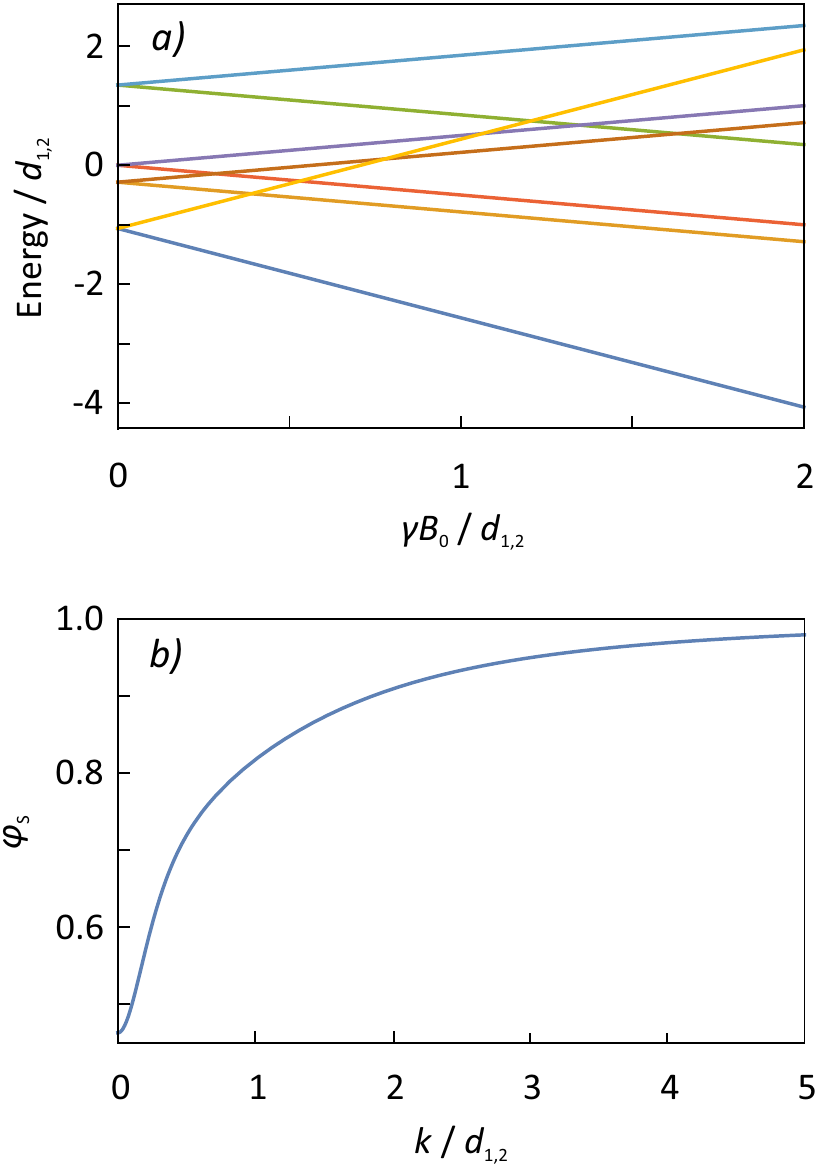}
 \caption{A linear spin triad with no exchange interaction. a) Energy level scheme for the field parallel to the
 molecular axis. b) Dependence of the singlet yield on the recombination rate constant $k$ for the magnetic field
 parallel to the $z$-axis (any field intensity).}
 \label{si1}
 \end{figure}

In the case with the magnetic field parallel to the symmetry axis, no MFEs are observed as discussed in the main text. The eigenstates and associated energies, which depend linearly on the field intensity (see Fig. 1) can be readily evaluated for this scenario. Further, the singlet yield can be calculated analytically following the approach as outlined in the main text. We obtain the following dependence upon the recombination rate constant $k$:

\begin{flalign}
 \varphi _S = \frac{{841}}{{1816}} + \frac{{147}}{{908}}L\left( {\frac{{\sqrt {681} }}{{16}}{d_{1,2}},k} \right) + \frac{{\left( {681 + 17\sqrt {681} } \right)}}{{3632}}L\left( {\frac{1}{{32}}\left( {\sqrt {681}  - 17} \right){d_{1,2}},k} \right) &\nonumber\\+ \frac{{\left( {681 - 17\sqrt {681} } \right)}}{{3632}}L\left( {\frac{1}{{32}}\left( {\sqrt {681}  + 17} \right){d_{1,2}},k} \right)
 \label{phi1}
\end{flalign}
where, $L(\Delta \omega ,k) = \frac{{{k^2}}}{{{k^2} + \Delta {\omega ^2}}}$.
Fig.~\ref{si1}(b) shows the dependence of yield on the recombination rate constant.

\begin{figure}
 \centering
  \includegraphics{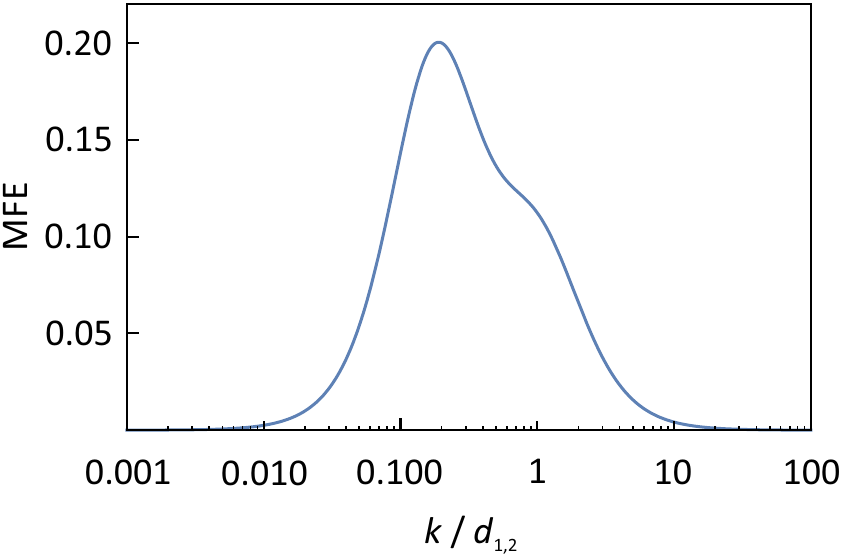}
 \caption{MFE of the linear spin-triad with the field perpendicular to the triad-axis as a function of the recombination rate constant $k$; no exchange; analytical result.}
 \label{si2}
 \end{figure}

\begin{figure}
\includegraphics{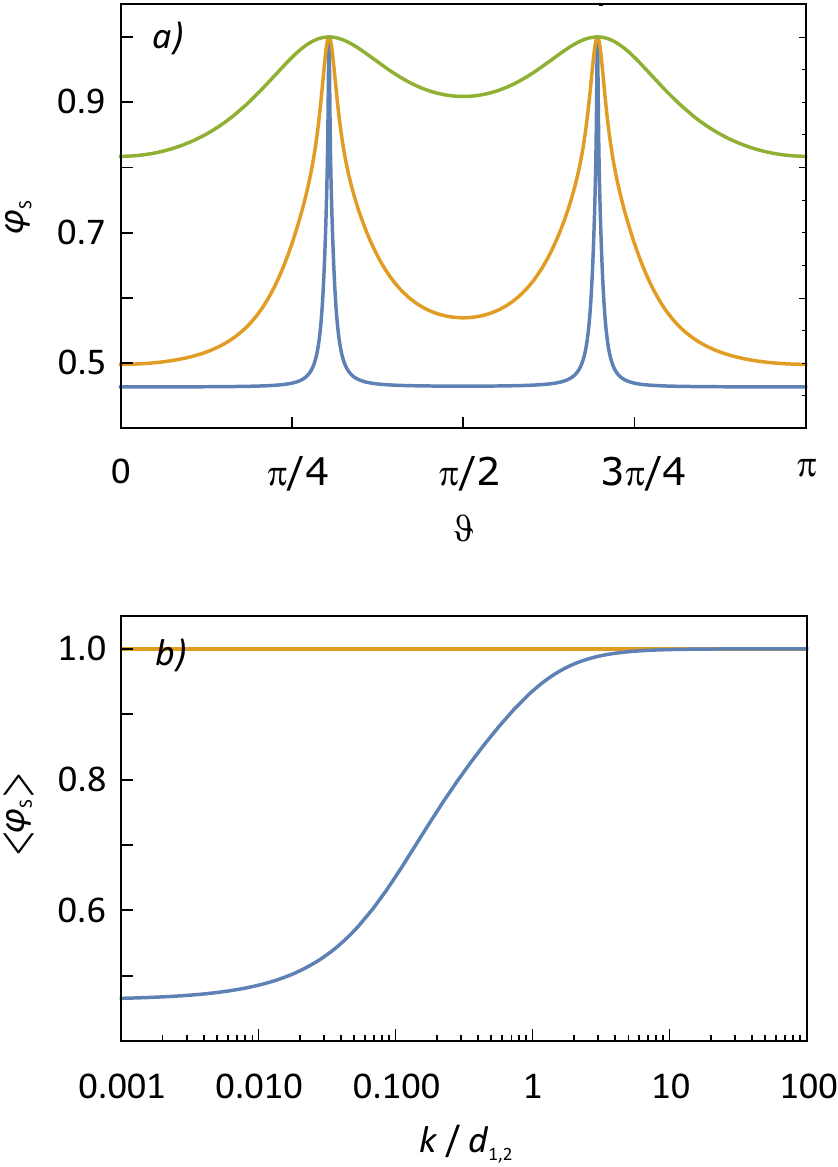}
\caption{Linear geometry with no exchange; analytical results. a) Directional dependence of the singlet yield, $k = \{0.01, 0.1, 1\}~d_{1,2}$. b) Powder-averaged singlet yield as a function of $k/d_{1,2}$.}
\label{si3}
\end{figure}

\begin{figure}
 \centering
  \includegraphics{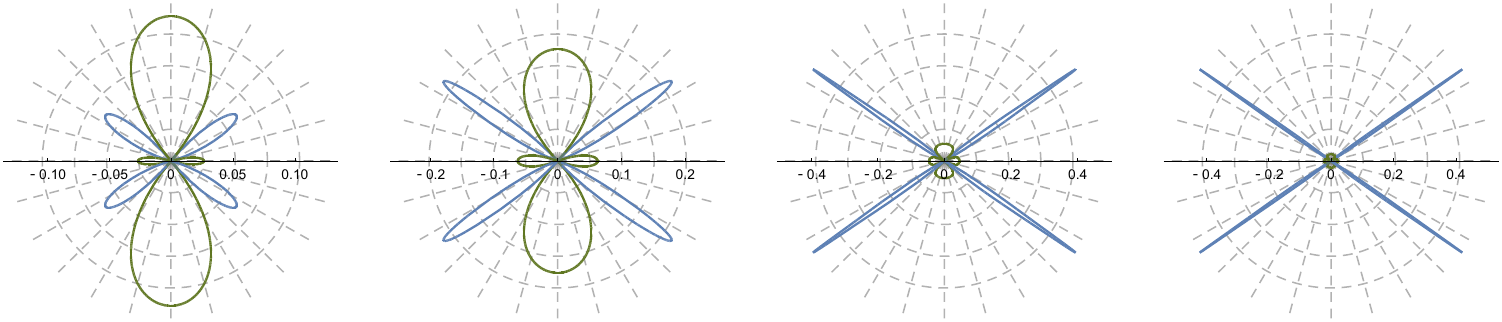}
 \caption{Polar plots of the anisotropy of the singlet yield for the linear geometry in the high-field limit. No exchange was used, and $k = \{0.01, 0.025, 0.25, 1\}d_{1,2}$.}
\label{si4}
 \end{figure}

For general orientations, an analytical calculation is impractical. An exception is the high-field scenario ($\gamma B_0 >> d_{1,2}$) with the magnetic field oriented perpendicular to the distinguished axis, which allows an approximate solution by degenerate perturbation theory. By representing the Hamiltonian in the eigenbasis of $\hat{H}_1$, followed by diagonalizing the degenerate sub-blocks, we obtain the following approximate expression for the singlet yield:
\begin{flalign}
\varphi _S \approx \frac{{841}}{{1816}} + \frac{{147}}{{908}}L\left( {\frac{{\sqrt {681} }}{{32}}{d_{1,2}},k} \right) + \frac{{\left( {681 + 17\sqrt {681} } \right)}}{{3632}}L\left( {\frac{1}{{64}}\left( {\sqrt {681}  - 17} \right){d_{1,2}},k} \right) &\nonumber\\ + \frac{{\left( {681 - 17\sqrt {681} } \right)}}{{3632}}L\left( {\frac{1}{{64}}\left( {\sqrt {681}  + 17} \right){d_{1,2}},k} \right).
\end{flalign}
Together with Eq. \ref{phi1} this allows one to approximately calculate the MFE for $B_0$ $>>$ $d_{1,2}$ and with the magnetic field at a perpendicular orientation. Fig. \ref{si2} illustrates this MFE as a function of the recombination rate constant. Remarkably, large MFEs can be realized for recombination rates of the order of $0.1$ to $1~d_{1,2}$, i.e. under conditions of relatively fast recombination.
In fact, this high-field procedure can be applied for an arbitrary orientation of the magnetic field. Denoting the angle between the axis of the three-spin system and the magnetic field by $\theta$, we obtain:
\begin{flalign}
 \varphi_S \approx \frac{{841}}{{1816}} + \frac{{147}}{{908}}L\left( {\frac{1}{{64}}\sqrt {681} \,h(\vartheta ){\kern 1pt} {d_{1,2}},k} \right) + \frac{{681 + 17\sqrt {681} }}{{3632}}L\left( {\frac{1}{{128}}\left( {\sqrt {681}  - 17} \right)\,h(\vartheta ){\kern 1pt} {d_{1,2}},k} \right) &\nonumber\\+ \frac{{681 - 17\sqrt {681} }}{{3632}}L\left( {\frac{1}{128}\left( {\sqrt {681}  + 17} \right)}h(\vartheta){\kern 1pt} {d_{1,2}, k} \right),
\end{flalign}
where $h(\vartheta) = 1 + 3\cos(2\vartheta)$.
It is noteworthy that the singlet yield is equal to $1$ for the magic angle, $\vartheta = \frac{1}{2}\arccos\left(-\frac{1}{3}\right)$, for which the secular terms of the dipole-dipole coupling (cf. below) vanish and thus no singlet-triplet mixing is induced. 
Fig. \ref{si3} shows the directional dependence of the singlet yield as given by Eq.~(16) and the dependence of the mean singlet yield on $k$. Fig. \ref{si4} illustrates the anisotropy of the singlet yield through polar plots of the deviation of the singlet yield from the spherical mean.  Note that remarkably spiky changes in the singlet yield can appear at the magic angle if the lifetime is long. 
For long lifetimes, $\varphi_S$ approaches $\frac{841}{1816}$; for fast recombination no spin mixing is realized and $\varphi_S$ approaches $1$.
Even in the presence of exchange interactions, no MFEs are observed when the field is along the symmetry axis. For this scenario, and $j_{1,2} = j_{2,3} = j~d_{1,2}$, and $j_{1,3} = 0$, the analytic dependence reads
\begin{flalign}
\varphi _S = \frac{1}{{s(j)}}\left[2\left({841 + 160j\left( {24j - 17} \right)} \right) + d(j)\left( {17 - 48j + d(j)} \right)L\left( {\frac{1}{{32}}\sqrt {{{\left( {d(j) - 17 - 16j} \right)}^2}} {d_{1,2}},k} \right)\right] &\nonumber\\ -\frac{1}{{s(j)}}\left[d(j)\left( {17 - 48j - d(j)} \right)L\left( {\frac{1}{{32}}\sqrt {{{\left( {d(j) - 17 - 16j} \right)}^2}} {d_{1,2}},k} \right) + 588\,L\left( {\frac{{d(j)}}{{16}}{d_{1,2}},k} \right)\right] 
\end{flalign}
where $s(j) = 16(227 +32 j(24j - 17))$ and $d(j) = \sqrt{3}\sqrt{227 - 544j + 768j^2}$.

\begin{figure}
\includegraphics{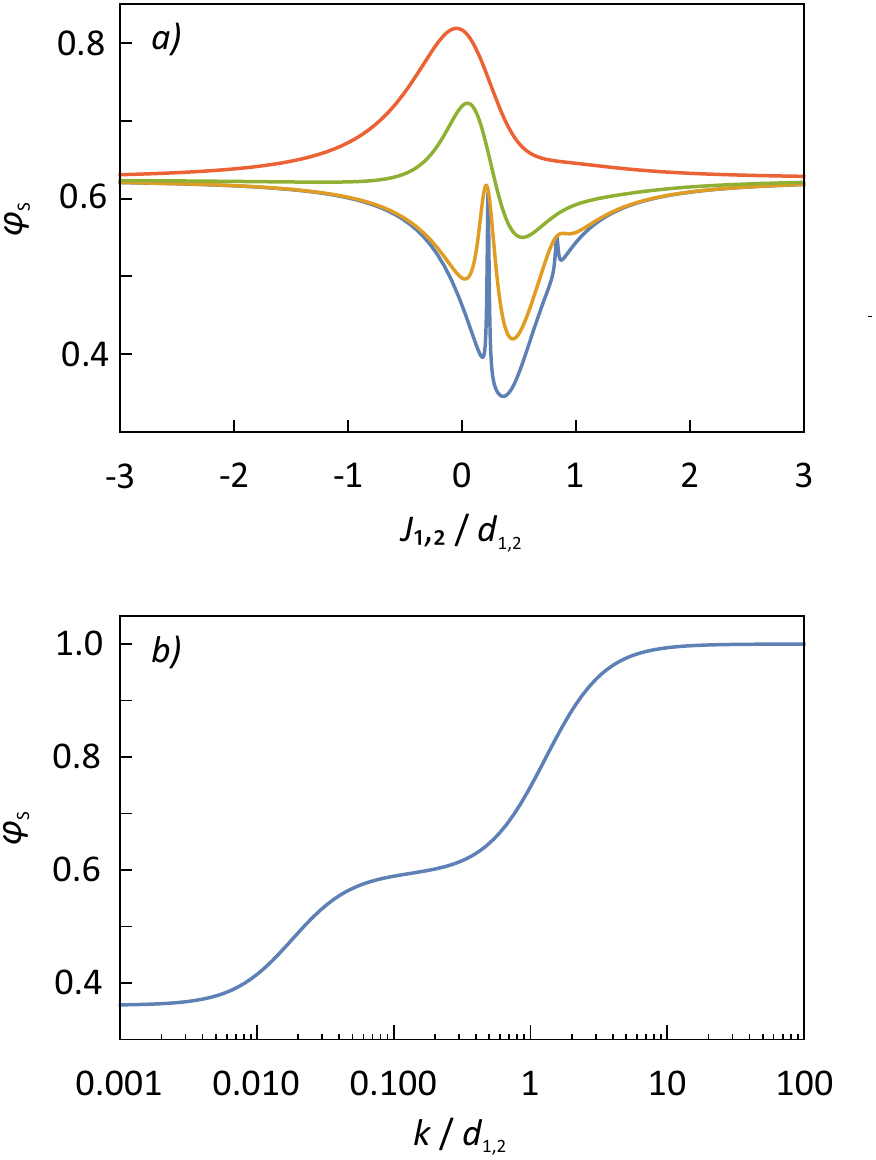}
\caption{Linear spin triad with exchange, $J_{1,2} = J_{2,3}$, and $J_{1,3} = 0$. a) Singlet yield as a function of $J_{1,2} = J_{2,3}$ for the magnetic field parallel to the molecular axis (any intensity). $k = \{0.01~\text{(blue)}, 0.1, 0.5, 1~\text{(red)}\}~d_{1,2}$. b) Dependence of the singlet yield on $k$ for $J_{1,2} = J_{2,3}$ = $\frac{1}{4}d_{1,2}$.}
\label{si5}
\end{figure}
\begin{figure}
\includegraphics{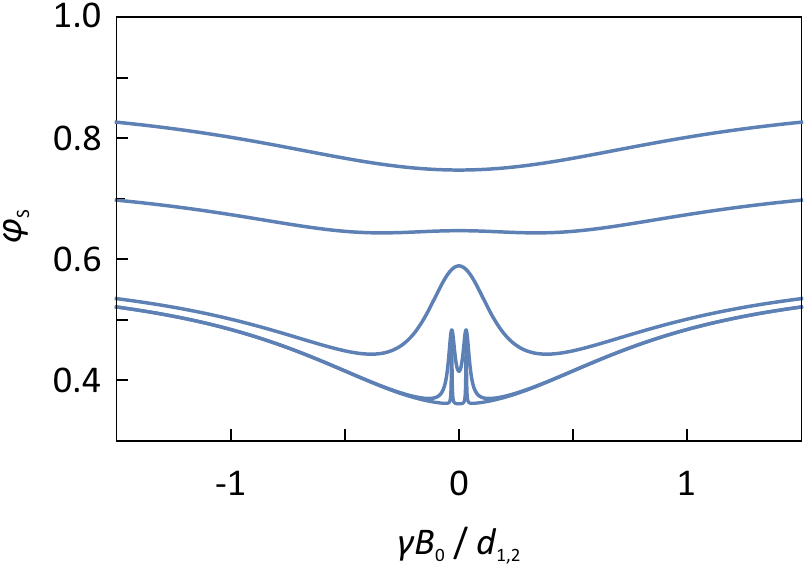}
\caption{Linear spin triad with $J_{1,2} = J_{2,3}$ = $\frac{1}{4}d_{1,2}$ and the field perpendicular to the molecular axis. ~~~~~~~~~~~~~~~~~~~~~~~~~~~~~~~~~~~~~~~~~~~~~~~~~~~~~~~~~~~~~~~~
$k = \{0.001, 0.01, 0.1, 0.5, 1\}~d_{1,2}$.}
\label{si6}
\end{figure}

A remarkable sensitivity to $k$ is recognized for $j$ in the range ($-1$,$1$), as is shown in Fig. \ref{si5}. Fig. \ref{si6} shows the yield as a function of magnetic field for different $k$ values when the magnetic field is along a perpendicular direction so that $[\hat{H}_0, \hat{H}_1] \neq 0$. It can be seen that the width of the low field peak broadens as $k$ increases.

\section{S6. On the origin of sharp spikes in the MFE of linear spin chains}

The spin Hamiltonian of linear spins chains with a transverse magnetic field can be approximately cast into the form of the $XXZ$ model, for which
\begin{equation}
\begin{aligned}
\hat{H}_{XXZ}&=\hat{H}_{0}+\hat{H}_{1}\\&= \sum_{i=1}^{N}[J_{i,i+1}\hat{S}_{i,x}\hat{S}_{j,x}+J_{i,i+1}\hat{S}_{i,y}\hat{S}_{j,y}+J_{i,i+1}\delta J_{i, i+1}\hat{S}_{i,z}\hat{S}_{j,z}]+\gamma B_{0}\sum_{i}^{N} \hat{S}_{i,x}
\end{aligned}
\end{equation}
where $J_{i, i+1}$ is a coupling parameter and $\delta J_{i, i+1}$ evaluates the anisotropy of this nearest-neighbour coupling. For $\delta J_{i, i+1} \neq 1$ the SU(2) symmetry is broken and the model resembles the dipole-dipole coupled spin chain. In particular, magnetic field effects are observed if the singlet projection operator does not commute with $\hat{H}_0$.

The sharp peak in reaction yield as a function of magnetic field is a consequence of crossings of eigenstates with different permutation symmetries. The linear spin model exhibits rich symmetries such as reflection symmetries and symmetries under global $\pi$ rotation around the $x$ axis. The presence of such symmetries may give rise to sharp peaks in the reaction yield for linear spin systems. For example, a linear four spin system described by $\hat{H}_{XXZ}$ has a global $\pi$ rotation symmetry about the $x$-axis, i.e $[\hat{H}_{XXZ},\hat{R}_{\pi}^{x}]=0$, where $\hat{R}_{\pi}^{x}=\hat{\sigma}_{x}^{1}\hat{\sigma}_{x}^{2}\hat{\sigma}_{x}^{3}\hat{\sigma}_{x}^{4}$. The eigenstate 
\begin{equation}
\left|\psi^{(i)}\right>=c_{1}^{i}\left|\alpha\alpha\beta\beta\right>+c_{2}^{i}\left|\alpha\beta\alpha\beta\right>+c_{3}^{i}\left|\alpha\beta\beta\alpha\right>+c_{4}^{i}\left|\beta\alpha\alpha\beta\right>+c_{5}^{i}\left|\beta\alpha\beta\alpha\right>+c_{6}^{i}\left|\beta\beta\alpha\alpha\right>
\end{equation}
has $c_{1}^{i}=c_{6}^{i}$, $c_{2}^{i}=c_{5}^{i}$, $c_{3}^{i}=c_{4}^{i}$, corresponding to an eigenvalue of $\hat{R}_{\pi}^{x}$ $E_{1}=+1$, or $c_{1}^{i}=-c_{6}^{i}$, $c_{2}^{i}=-c_{5}^{i}$, $c_{3}^{i}=-c_{4}^{i}$, corresponding to an eigenvalue of $\hat{R}_{\pi}^{x}$ $E_{2}=-1$.  
Furthermore, we can define a parity operator for linear spin systems as 
\begin{equation}
\hat{P}={\hat{P}_{1,N}\hat{P}_{2,N-1}...\hat{P}_{\frac{N}{2},\frac{N+2}{2}}}
\end{equation}
for even $N$, and 
\begin{equation}
\hat{P}={\hat{P}_{1,N}\hat{P}_{2,N-1}...\hat{P}_{\frac{N-1}{2},\frac{N+3}{2}}}
\end{equation}
for odd $N$. If $[\hat{H}, \hat{P}]=0$, then the parity is conserved and the eigenstates of the Hamiltonian can be classified as even or odd. For example, an eigenstate $\left|\psi^{(i)}\right>=c_{1}^{i}\left|\alpha\beta\beta\beta\right>+c_{2}^{i}\left|\beta\alpha\beta\beta\right>+c_{3}^{i}\left|\beta\beta\alpha\beta\right>+c_{4}^{i}\left|\beta\beta\beta\alpha\right>$ the even parity gives $c_{1}^{i}=c_{4}^{i}$, $c_{2}^{i}=c_{3}^{i}$ and odd parity gives $c_{1}^{i}=-c_{4}^{i}$, $c_{2}^{i}=-c_{3}^{i}$. 
If an eigenstate of a certain symmetry (e.g. with respect to parity or the global $x$-rotation), $\left| i\right>$ undergoes crossing with another eigenstate $\left| j\right>$ with different symmetry properties such that $\left<i\right|\hat{P}_{s}^{(i,j)}\left|j\right>\neq0$, the sharp spikes occur in the reaction yield. It is indeed interesting that such an interplay of symmetries, level crossings and many-body interactions can lead to pronounced MFEs at fields lower than the geomagnetic field.


\section{S7. MFEs in the equilateral triangular geometry}

We further investigate the MFEs in an equilateral triangular geometry. We choose the position vectors of the spins as
$\roarrow{\mathbf{r}}_{1}= r(\frac{1}{2};0;0)$, $\roarrow{\mathbf{r}}_{2}= r(-\frac{1}{2};0;0)$ and $\roarrow{\mathbf{r}}_{3}= r((0, \sin(60$\textdegree));$0;0)$. The corresponding dipolar tensors are of the form

\begin{equation} D_{1,2}=
\begin{pmatrix}
-2 &0 &0\\
0 & 1 & 0\\
0 & 0 & 1
\end{pmatrix}d_{1,2}
\end{equation}

\begin{equation} D_{2,3}=
\begin{pmatrix}
\frac{1/4} & \frac{-3\sqrt{3}}{4} & 0\\
\frac{-3\sqrt{3}}{4} & -\frac{5}{4} & 0\\
0 & 0 & 1
\end{pmatrix}d_{1,2}
\end{equation}

\begin{equation} D_{1,3}=
\begin{pmatrix}
\frac{1/4} & \frac{3\sqrt{3}}{4} & 0\\
\frac{3\sqrt{3}}{4} & -\frac{5}{4} & 0\\
0 & 0 & 1
\end{pmatrix}d_{1,2}
\end{equation}

For $B_0 = 0$, the singlet yield can be evaluated analytically. We obtain:
\begin{equation}
\varphi _S = \frac{5}{{13}} + \frac{3}{{26}}L\left( {\frac{{3\sqrt {13} }}{4}{d_{1,2}},k} \right) + \frac{{13 + \sqrt {13} }}{{52}}L\left( {\frac{3}{8}\left( {\sqrt {13}  - 1} \right){d_{1,2}},k} \right) + \frac{{13 - \sqrt {13} }}{{52}}L\left( {\frac{3}{8}\left( {\sqrt {13}  + 1} \right){d_{1,2}},k} \right)\
\end{equation}

In addition, for the field along the $z$-direction, the following expression for the singlet yield can be derived

\begin{flalign}
\varphi _S = \frac{1}{4}\left( 4 - \frac{{27}}{{117 + 16{r^2} + 16b\left( {4b - 3} \right)}} - \frac{{27}}{{117 + 16{r^2} + 16b\left( {4b + 3} \right)}}\right) &\\\nonumber - \frac{1}{4}\left(\frac{{27\left( {16{r^2} + 27} \right)}}{{729 + 16{r^2}\left( {63 + 16{r^2} + 16b\left( {4b - 3} \right)} \right)}} - \frac{{27\left( {16{r^2} + 27} \right)}}{{729 + 16{r^2}\left( {63 + 16{r^2} + 16b\left( {4b + 3} \right)} \right)}} \right).
\end{flalign}
Here, $b =  \frac{\gamma B_0}{d_{1,2}}$ and $r = \frac{k}{d_{1,2}}$. Note that $\varphi_S$ approaches $1$ as $B_0 \to \infty$.


\section{S8. Blocked singlet-triplet conversion in the high-field limit}

In the high-field limit, the quantisation axis of the spins corresponds to the direction of the magnetic field. It is then possible to simplify the Hamiltonian $\hat{H}_0$ to only include the secular terms:

\begin{equation}
\hat{H}'_{dd} = \frac{1}{2}\sum\limits_{i < j} {{d_{i,j}}({r_{i,j}})} (3\cos {\theta _{i,j}} - 1)\left( {3{{\hat S}_{i,z}}{{\hat S}_{j,z}} - {{{\mathbf{\hat S}}}_i} \cdot {{{\mathbf{\hat S}}}_j}} \right)
\end{equation}
where $\theta_{i,j}$ is the angle between the magnetic field and the vector connecting spins $i$ and $j$. The individual terms vanish for the magic angle $\theta_{i,j} = 54.7$\textdegree, which for the linear configuration can be used to simultaneously suppress all dipolar couplings, i.e. $\hat{H}'_{dd} = 0$ and thus the singlet-triplet mixing vanishes (\textit{vide supra}). For general geometries it is not possible to simultaneously fulfil the three magic-angle conditions. Nonetheless, the more broad condition $[\hat{P}_{S}^{(1,2)}, \hat{H}'_{dd}] = 0$ can be realized for certain field orientations for which consequently $\varphi_S = 1$. Assuming an isosceles triangular arrangement with bond angle $\alpha$ it can be shown that $\hat{H}'_{dd}$ and $\hat{P}_{S}^{(1,2)}$ commute if the following condition is satisfied:

\begin{equation}
\begin{gathered}
  \resizebox{0.98\hsize}{!}{$\frac{1}{8}{\csc ^3}\left( {\tfrac{\alpha }{2}} \right)\left( \left( {1 + 3\cos \left( {2\vartheta } \right)} \right)\left( {2 - 6\sin \left( {\tfrac{\alpha }{2}} \right) + 7\sin \left( {\tfrac{{3\alpha }}{2}} \right) - 3\sin \left( {\tfrac{{5\alpha }}{2}} \right)} \right) + 6\cos \left( {2\phi } \right)\left( {10\sin \left( {\tfrac{\alpha }{2}} \right) - 5\sin \left( {\tfrac{{3\alpha }}{2}} \right) + \sin \left( {\tfrac{{5\alpha }}{2}} \right) - 2} \right){{\sin }^2}\left( \vartheta  \right) \right) \hfill$} \\
  \quad  + 12\cos \left( \phi  \right)\sin \left( \alpha  \right)\sin \left( {2\vartheta } \right) = 0 \hfill \\
\end{gathered} 
\label{contours}
\end{equation}

Here, the spin-triad has been placed in the $x$,$z$-plane ($z \leq 0$) and $\phi$ and $\theta$ are the polar and azimuthal angle specifying the direction of the magnetic field vector. Typically, Eq. \ref{contours} defines $2$ closed contours on the $\theta,\phi$-sphere. For the linear configuration ($\alpha = \pi$), Eq. \ref{contours} is tantamount to the vanishing of the three dipolar couplings. For $\alpha < \pi$, the dipolar coupling gives rise to large anisotropies of the MFE, which typically exceed the anisotropies from hyperfine-induced singlet-triplet mixing, because the latter lacks this unique switching property.


\section{S9. MFEs in more general geometries}

Fig. \ref{BondAngle} illustrates the bond angle dependence of the MFE for isosceles triangular spin triads. Fig. \ref{Density_BisGeo} and Fig. \ref{Density_BisD} show the absolute value and anisotropy of the singlet yield, for more general triad geometries together with the absolute value of the average MFE and its maximal absolute MFE value in a particular instance.

Figs. \ref{LinEx} to \ref{Isosceles} show the powder averages of the singlet yield for dipolarly-coupled triads in the linear, equilateral triangular and isosceles geometries, respectively. They also depict, for each case, the field of half-saturation, locations of low-field minima and maxima, and absolute MFE values all as a function of the recombination rate constant $k$.

\begin{figure}
\includegraphics{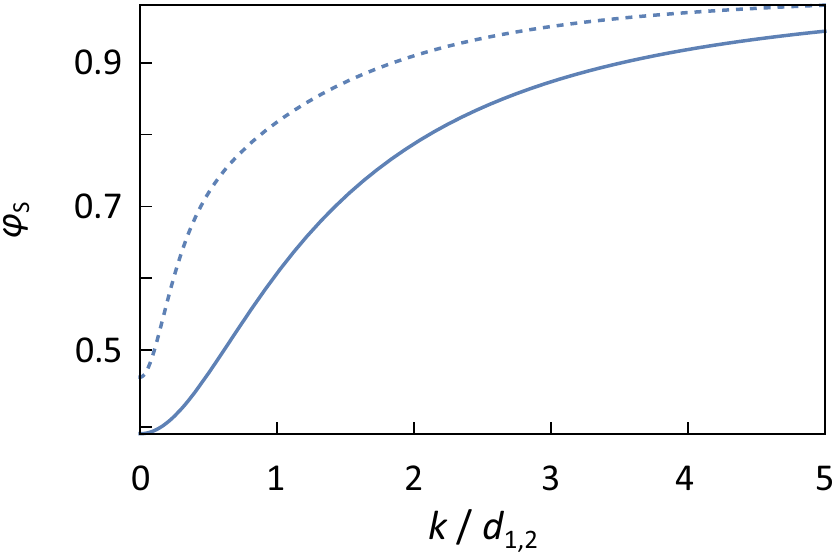}
\caption{Singlet yield for $B_0 = 0$ for an equilateral triangular geometry (solid line) and the linear spin triad (dashed line) as a function of the recombination rate constant $k$.}
\label{si7}
\end{figure}

\begin{figure}
\includegraphics{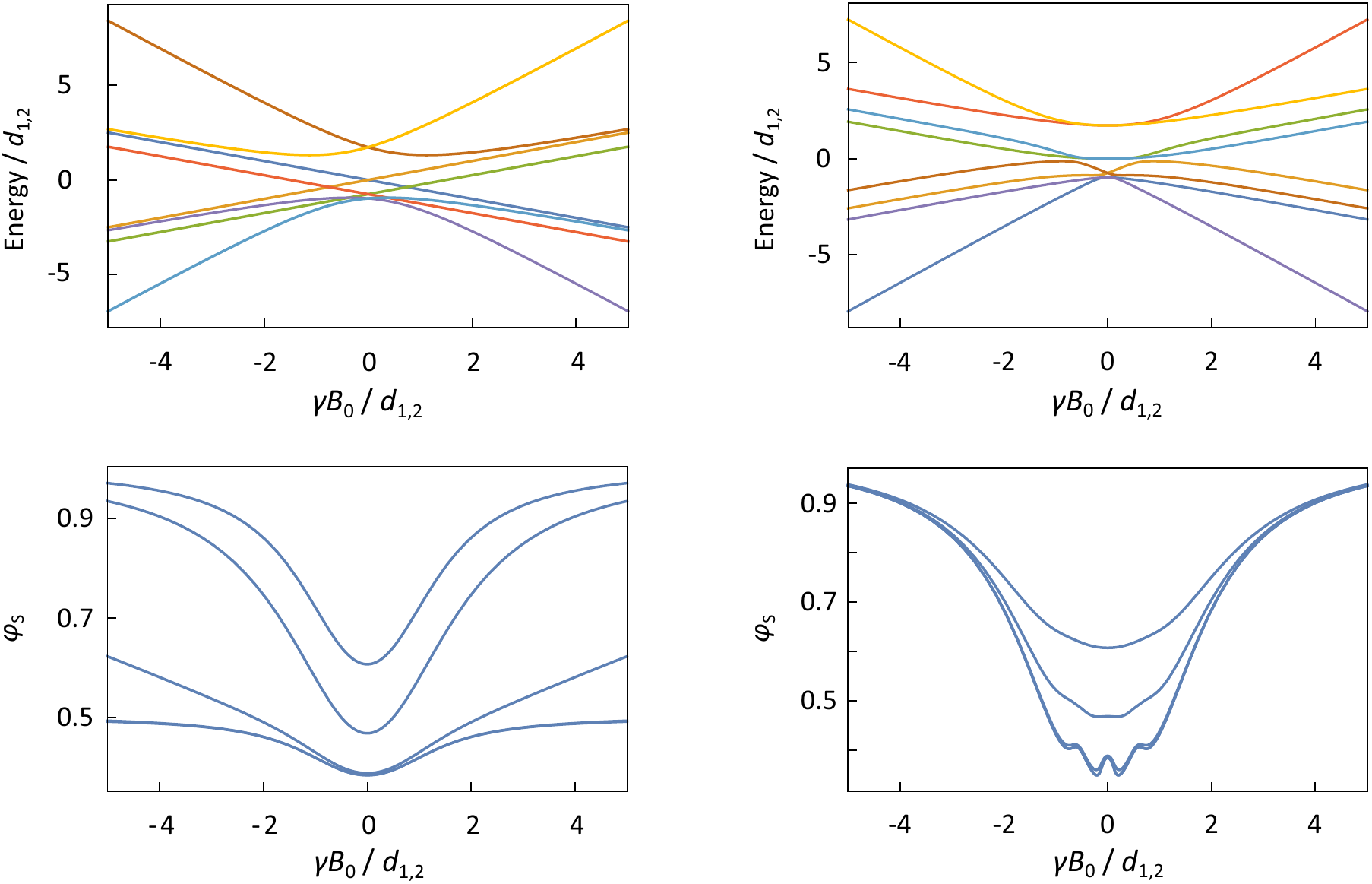}
\caption{Energies of the eigenstates (top) and singlet yields (bottom) as a function of the magnetic field for a spin triad with an equilateral triangular geometry and the field oriented perpendicular to the ring plane ($z$-axis; left) or along the $x$- or $y$-axis (right). The singlet yields are plotted for $k = \{0.001, 0.01, 0.1, 0.5, 1\}~d_{1,2}$, with the lowest $k$ corresponding to the bottommost curve; the curves for $k = 0.001~d_{1,2}$ and  $k = 0.01~d_{1,2}$ practically coincide.}
\label{si8}
\end{figure}

\begin{figure}
\includegraphics{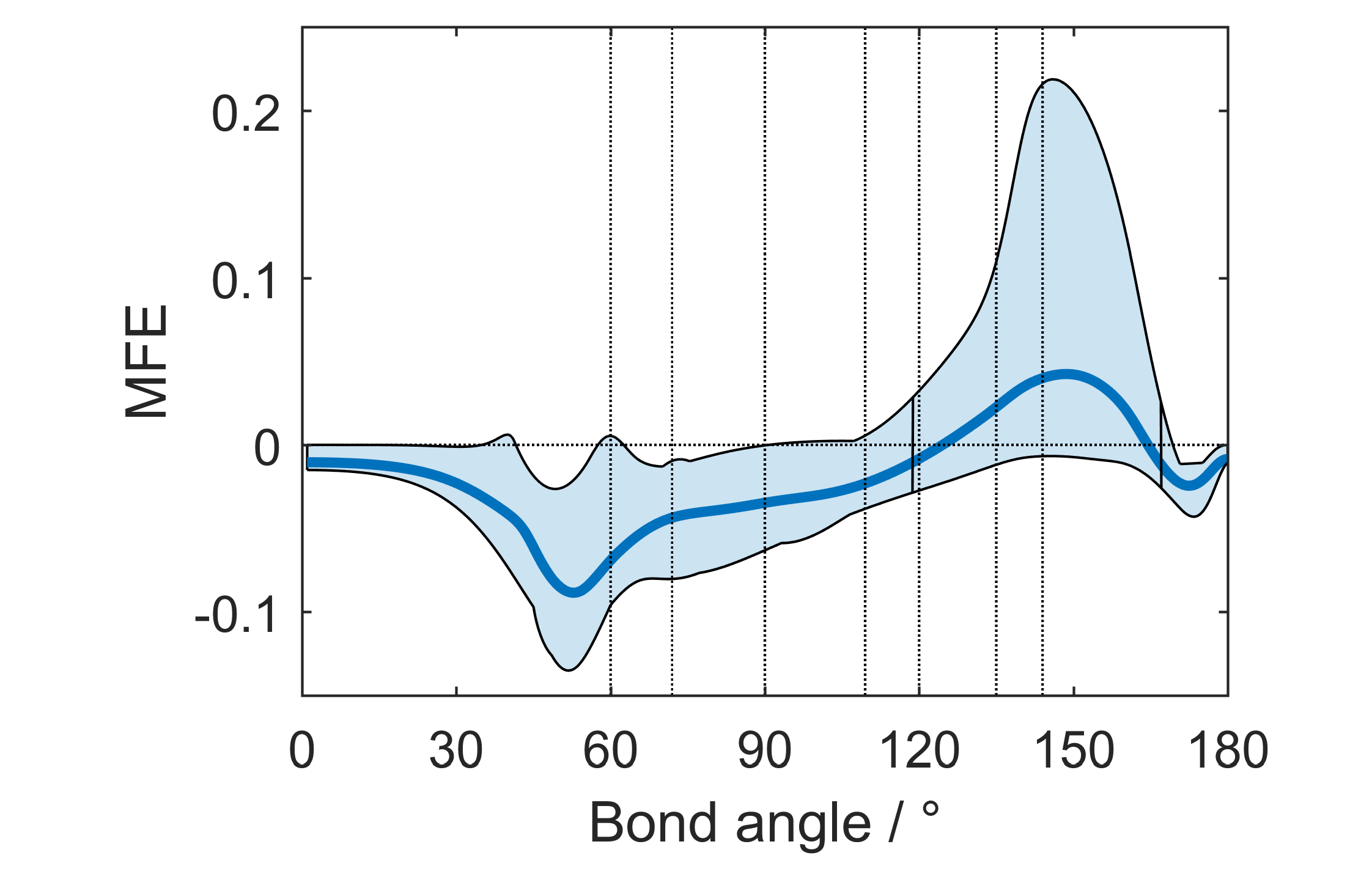}
\includegraphics{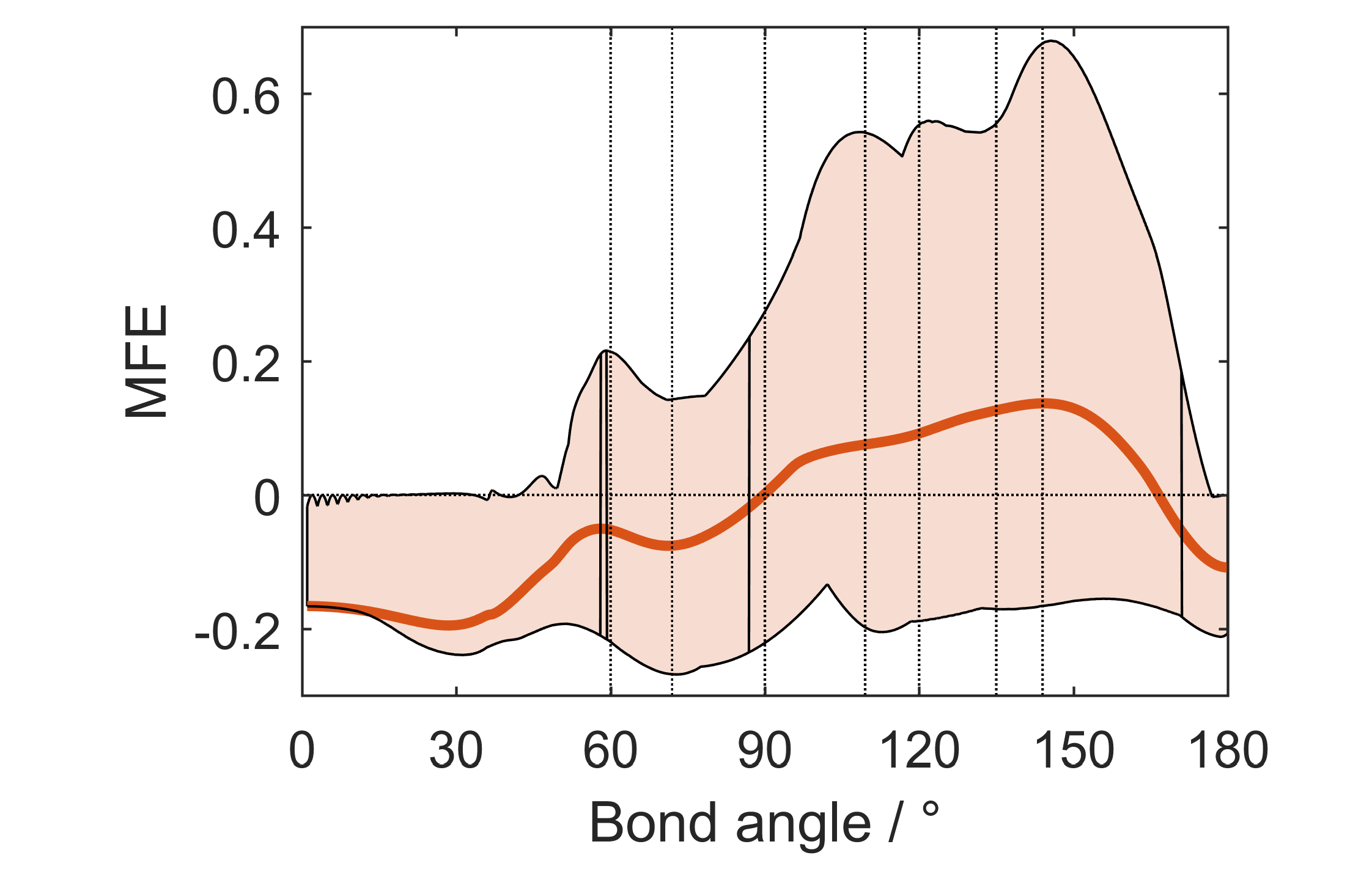}
\includegraphics{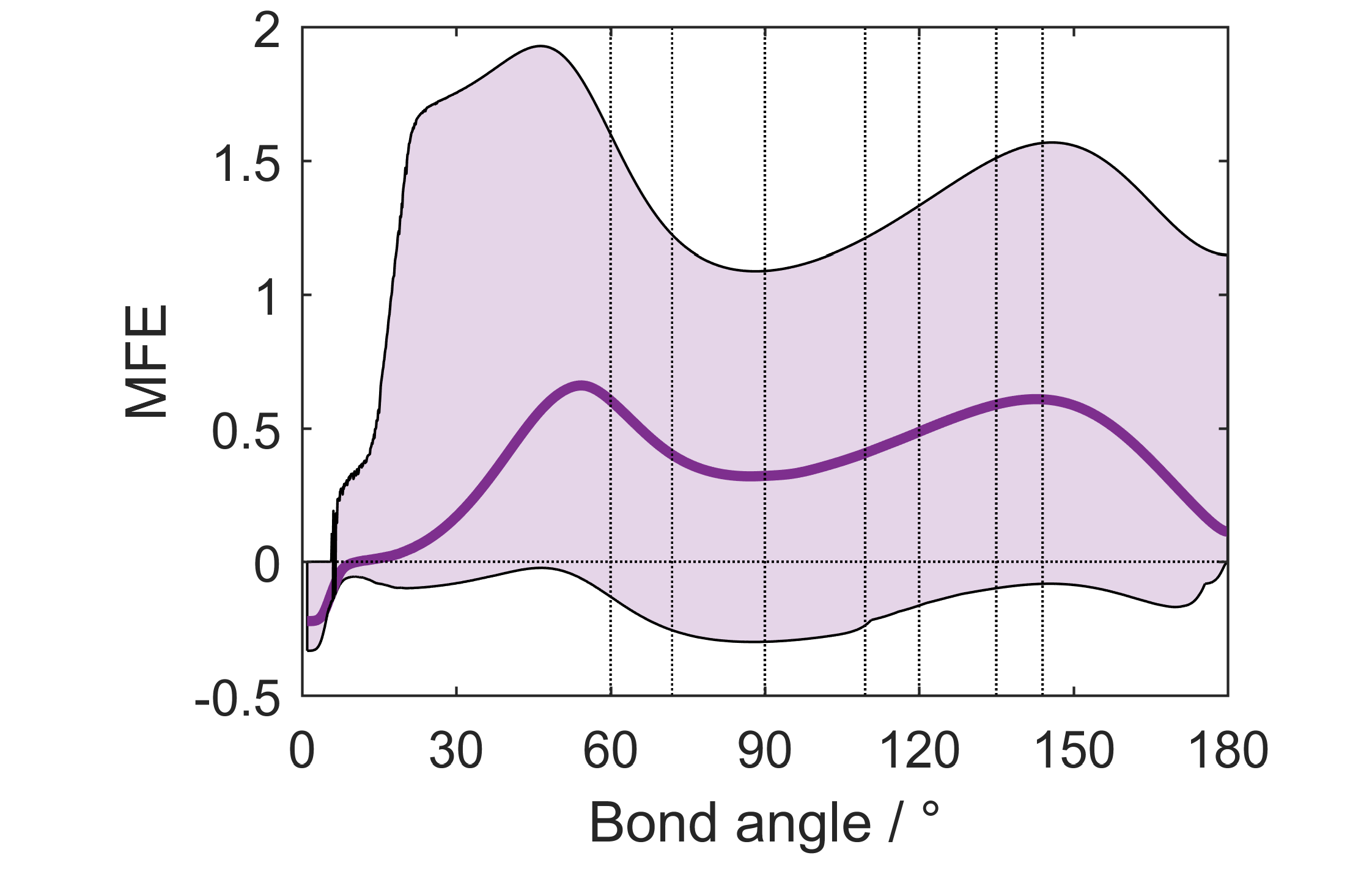}
\caption{Bond angle dependence of the MFE for isosceles triangular spin triads with an adjacent radical distance of $20~$\AA~and spin $2$ at the centre for the Earth's magnetic field ($B_0 = 50~\mu T$; top), $\gamma B_0 = d_{1,2}$ (centre) and $\gamma B_0 = 1000~d_{1,2}$ (bottom). The plots show the average of the MFE for an ensemble of randomly oriented spin triads and the orientational spread of the MFE. The latter is a measure of the directional anisotropy of the effect at the given field intensity. A lifetime of $k^{-1} = 1~\mu s$ was assumed throughout.}
\label{BondAngle}
\end{figure}

\begin{figure}
 \centering
  \includegraphics{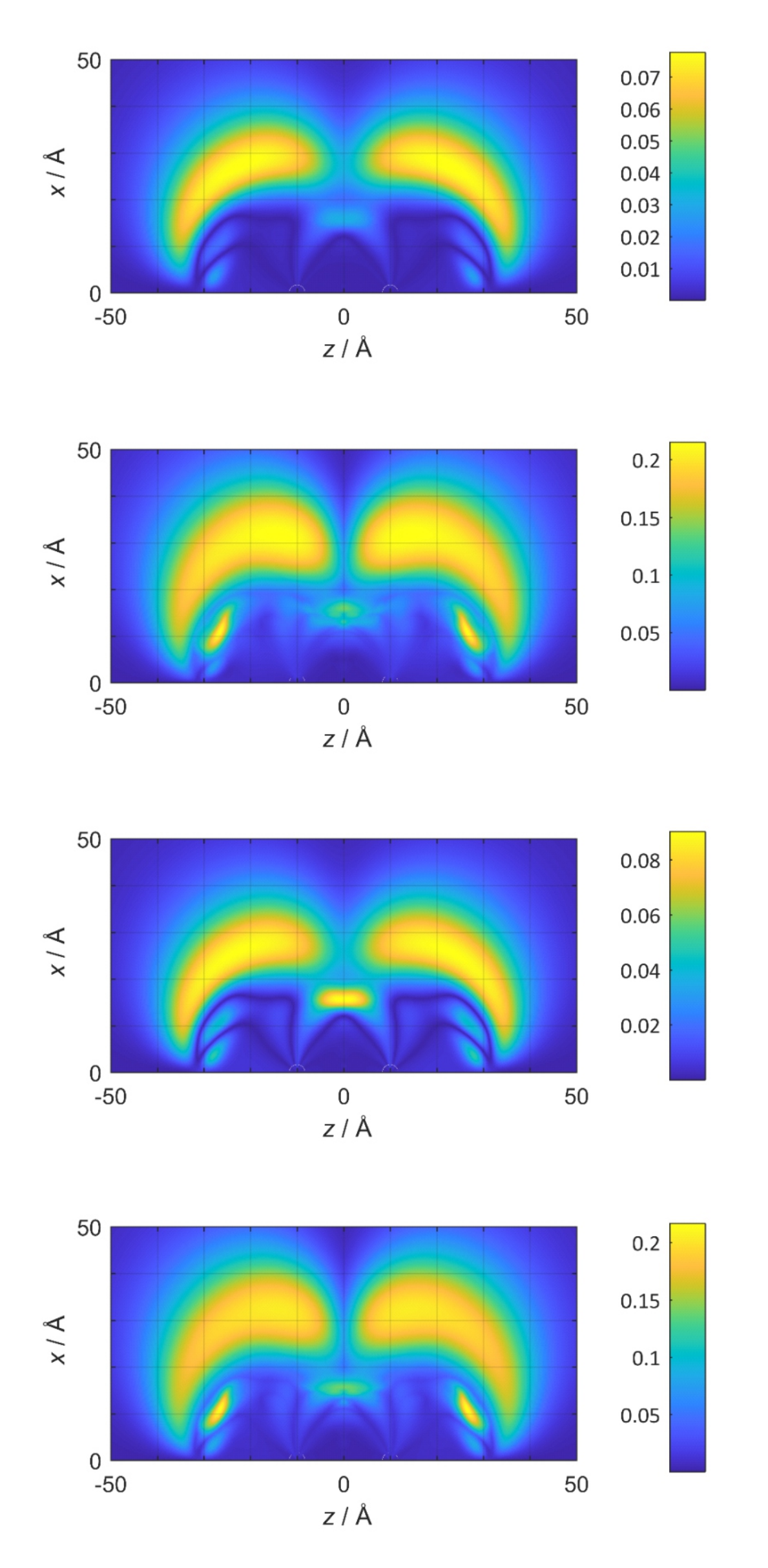}
 \caption{Three spins, various configurations. One spin is at $(0, -10)$~\AA, one at  $(0, 10)$~\AA; the location of the third spin is varied in the ($x$,$z$)-plane; $B_0 = 50~\mu T$, $k^{-1} = 1~\mu s$. Top: Absolute value of the change in the orientationally averaged singlet yield, anisotropy of the singlet yield (i.e. the maximal absolute change divided by the mean effect), absolute value of the average MFE, and the maximal absolute value of the MFE realized for a particular orientation.}
 \label{Density_BisGeo}
 \end{figure}

\begin{figure}
 \centering
  \includegraphics{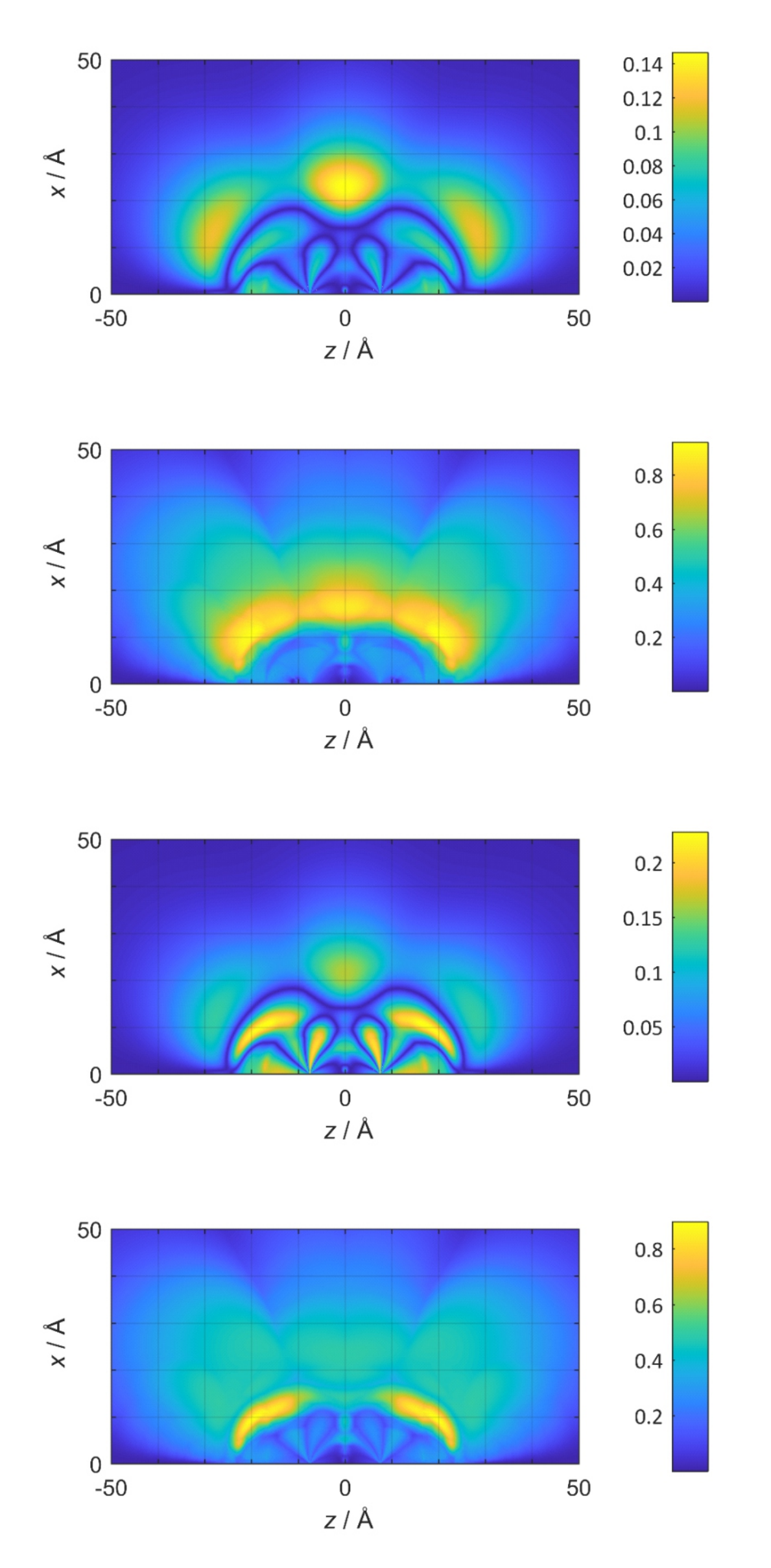}
 \caption{Three spins, various configurations. One spin is at $(0, -7.5)$~\AA, one at  $(0, 7.5)$~\AA; the location of the third spin is varied in the ($x$,$z$)-plane; $B_0 = d_{1,2}$, $k^{-1} = 1~\mu s$. Top: Absolute value of the change in the orientationally averaged singlet yield, anisotropy of the singlet yield (i.e. the maximal absolute change divided by the mean effect), absolute value of the average MFE, and the maximal absolute value of the MFE realized for a particular orientation.}
 \label{Density_BisD}
 \end{figure}
 
 \begin{figure}
 \centering
  \includegraphics[height=13cm]{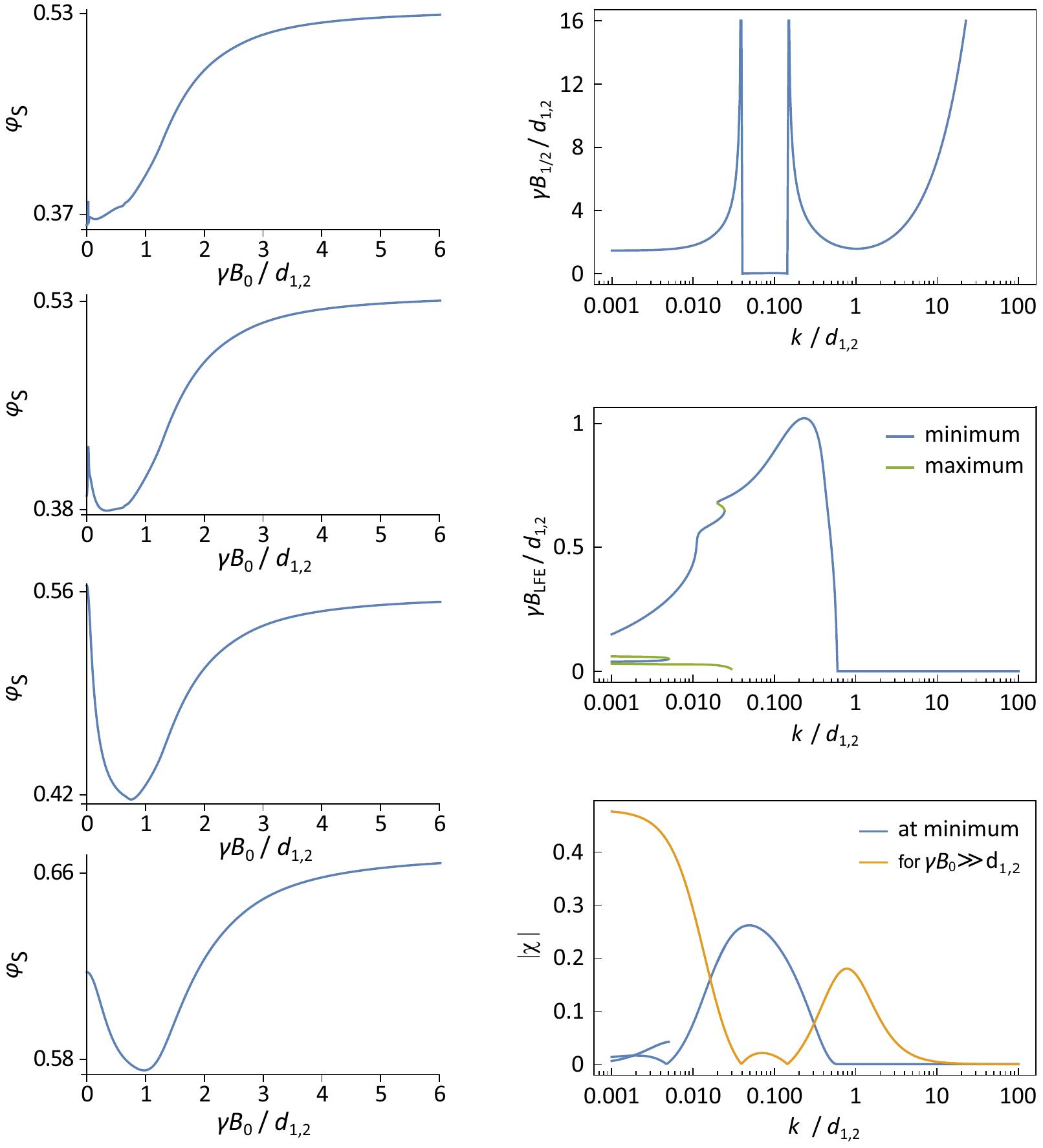}
 \caption{Powder averages of the singlet yield for a linear spin-triad, with all spins subject to electron-electron dipolar coupling and adjacent spins additionally subjected to exchange coupling: $J = J_{1,2} = J_{2,3} = 0.25~d_{1,2}$. Left: Dependence of the singlet yield on the magnetic field intensity for different recombination rate constants $k = [0.001~\text{(top)}, 0.00686, 0.0470, 0.322~\text{(bottom)}]~d_{1,2}.$ Right: Field of half-saturation (top), location of (low-field) minima and maxima of the field-dependence of the singlet yield (centre) and absolute values of MFEs for a saturating magnetic field and at the characteristic minima (bottom), all represented as a function of the recombination rate constant $k$.}
 \label{LinEx}
 \end{figure}

\begin{figure}
 \centering
  \includegraphics{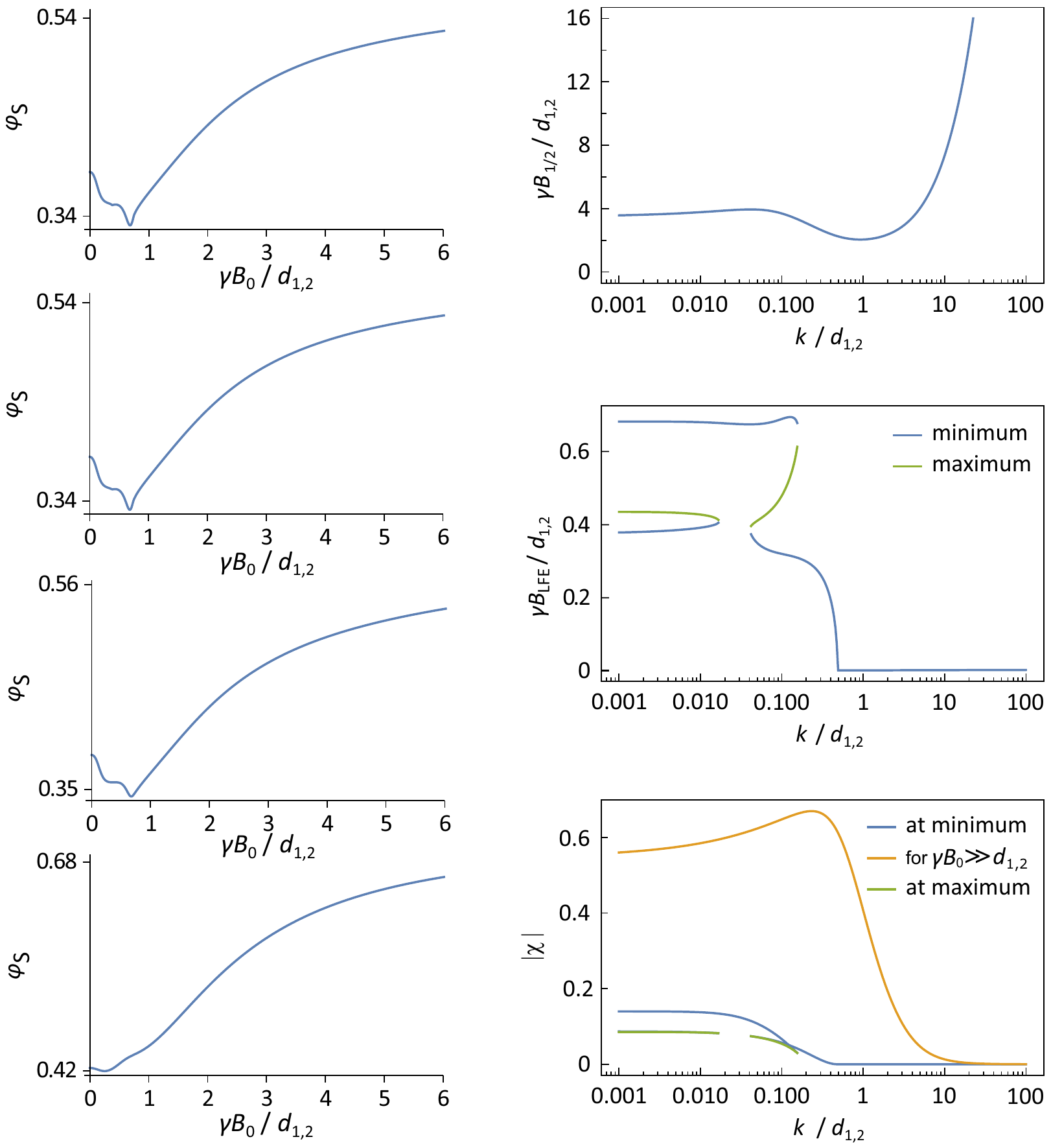}
 \caption{Powder averages of the singlet yield for a dipolarly coupled spin-triad in an equilateral triangular configuration. Left: Dependence of the singlet yield on the magnetic field intensity for different recombination rate constants $k = [0.001~\text{(top)}, 0.00686, 0.0470, 0.322~\text{(bottom)}]~d_{1,2}.$ Right: Field of half-saturation (top), location of (low-field) minima and maxima of the field-dependence of the singlet yield (centre), and absolute value of the MFEs for a saturating magnetic field and at the characteristic minima and maxima (bottom), all represented as a function of the recombination rate constant $k$.}
\label{Equi}
 \end{figure}

\begin{figure}
 \centering
  \includegraphics{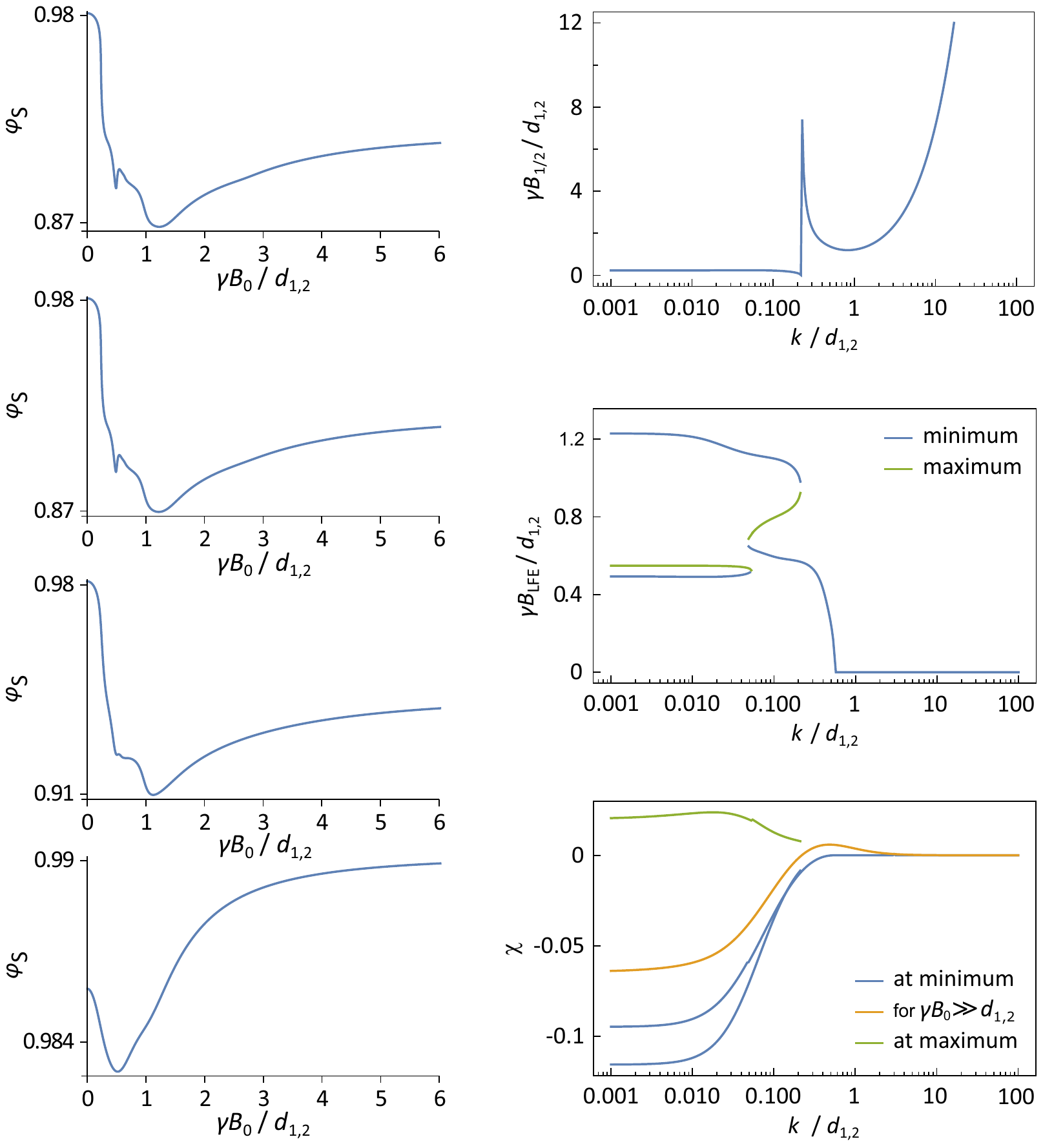}
 \caption{Powder averages of the singlet yield for a dipolarly coupled spin-triad in an isosceles triangular configuration, with the third radical at a distance of $2~r_{1,2}$ from radicals $1$ and $2$. Left: Dependence of the singlet yield on the magnetic field intensity for different recombination rate constants $k = [0.001~\text{(top)}, 0.00686, 0.0470, 0.322~\text{(bottom)}]~d_{1,2}.$ Right: Field of half-saturation (top), location of (low-field) minima and maxima of the field-dependence of the singlet yield (centre), and absolute value of the MFEs for a saturating magnetic field and at the characteristic minima and maxima (bottom), all represented as a function of the recombination rate constant $k$.}
\label{Isosceles}
 \end{figure}

\end{document}